\documentclass[10pt,journal,cspaper,compsoc]{IEEEtran}
\usepackage[cmex10]{amsmath}
\usepackage{algorithmic}
\usepackage{array}
\usepackage{mdwmath}
\usepackage{mdwtab}
\usepackage{eqparbox}
\usepackage{cite}
\usepackage{amssymb}
\usepackage{float}
\usepackage{eqparbox}
\usepackage{multirow}

\usepackage{algorithm}
\usepackage{algorithmic}
\usepackage{color}

\usepackage{array}

\usepackage{perpage}                  
\MakePerPage{footnote}                 

\usepackage{mdwmath}
\usepackage{mdwtab}
\usepackage{amsmath} \usepackage{amsfonts}
\makeatletter \def\amsbb{\use@mathgroup \M@U \symAMSb} \makeatother
\usepackage{bbold}
\usepackage{subfigure}
\usepackage{caption}
\usepackage{cite}
\usepackage{graphicx}
\usepackage{booktabs} 
\usepackage{lipsum}
\usepackage{amssymb}
\usepackage{amsthm}
\usepackage{eqparbox}
\usepackage{mathtools}
\usepackage{multirow}
\usepackage[cmex10]{amsmath}
\usepackage{stfloats}  

\usepackage{amsmath}
\usepackage{etoolbox}
\newcounter{numrel}
\renewcommand{\thenumrel}{\roman{numrel}}
\let\textlabel\label
\newcommand{\numrel}[2]{
  \begingroup%
  \refstepcounter{numrel}
  \textlabel{#2}\endgroup
  \ensuremath{\stackrel{(\thenumrel)}{#1}}
}
\AfterEndEnvironment{align}{\setcounter{numrel}{0}}
\usepackage{empheq}

\begin{document}
%
\title{Diffusion Adaptation over Multi-Agent Networks with Wireless Link Impairments}
\author{Reza~Abdolee,~\IEEEmembership{Student Member,~IEEE,}
        Benoit~Champagne,~\IEEEmembership{Senior Member,~IEEE,}
        and~ \mbox{Ali~H. Sayed},~\IEEEmembership{Fellow,~IEEE}%
\IEEEcompsocitemizethanks{\IEEEcompsocthanksitem{R. Abdolee and B. Champagne are with the Department
of Electrical and Computer Engineering, McGill University, Montreal,
QC, H3A 0E9 Canada (e-mail: reza.abdolee@mail.mcgill.ca, benoit.champagne@mcgill.ca). The work of B. Champagne and R. Abdolee was supported in part by an NSERC Discovery grant.}%
\IEEEcompsocthanksitem{A. H. Sayed is with the Department of Electrical Engineering, University
of California, Los Angeles, CA 90095 USA (e-mail:
sayed@ee.ucla.edu). The work of A. H. Sayed was supported in part by NSF CCF-1011918 and ECCS-1407712.}
}}
\def\H{\mbox{\boldmath $H$}}
\def\A{\mbox{\boldmath $A$}}
\def\D{\mbox{\boldmath $D$}}
\def\N{\mbox{\boldmath $N$}}
\def\Y{\mbox{\boldmath $Y$}}
\def\bphi{\mbox{\boldmath $\phi$}}
\def\bpsi{\mbox{\boldmath $\psi$}}
\def\btheta{\mbox{\boldmath $\theta$}}
\def\bbeta{\mbox{\boldmath $\beta$}}
\def\blambda{\mbox{\boldmath $\lambda$}}
\def\balpha{\mbox{\boldmath $\alpha$}}
\def\bgamma{\mbox{\boldmath $\gamma$}}
\def\bPsi{\mbox{\boldmath $\Psi$}}
\def\bomega{\mbox{\boldmath $\omega$}}
\def\bsigma{\mbox{\boldmath $\sigma$}}
\def\E{\amsbb{E}}
\def\diag{\mbox{\rm{diag}}}
\def\col{\mbox{\rm{col}}}
\def\conj{\mbox{\rm{conj}}}
\def\msd{{\rm{msd}}}
\def\emse{{\rm{emse}}}
\def\for{{\;\rm{for}\;}}
\def\bvec{{\rm{bvec}}}
\def\vec{{\rm{vec}}}
\def\var{{\rm{var}}}
\def\Ind{{\rm{Ind}}}
\def\rank{{\rm{rank}}}
\def\net{{\rm{net}}}
\def\Tr{{\rm{Tr}}}
\def\n{\boldsymbol{n}}
\def\g{\boldsymbol{g}}
\def\Ical{\boldsymbol{\cal I}}
\def\a{\boldsymbol{a}}
\def\e{\boldsymbol{e}}
\def\f{\boldsymbol{f}}
\def\h{\boldsymbol{h}}
\def\d{\boldsymbol{d}}
\def\p{\boldsymbol{p}}
\def\s{\boldsymbol{s}}
\def\t{\boldsymbol{t}}
\def\u{\boldsymbol{u}}
\def\v{\boldsymbol{v}}
\def\w{\boldsymbol{w}}
\def\x{\boldsymbol{x}}
\def\y{\boldsymbol{y}}
\def\z{\boldsymbol{z}}

\def\be{\begin{equation}}
\def\ee{\end{equation}}
\def\ba{\begin{align}}
\def\ea{\end{align}}

\IEEEcompsoctitleabstractindextext{%
\begin{abstract}
We study the performance of diffusion least-mean-square algorithms for distributed parameter estimation in multi-agent networks when nodes exchange information over wireless communication links. Wireless channel impairments, such as fading and path-loss, adversely affect the exchanged data and cause instability and performance degradation if left unattended. To mitigate these effects, we incorporate equalization coefficients into the diffusion combination step and update the combination weights dynamically in the face of randomly changing neighborhoods due to fading conditions. When channel state information (CSI) is unavailable, we determine the equalization factors from pilot-aided channel coefficient estimates. The analysis reveals that by properly monitoring the CSI over the network and choosing sufficiently small adaptation step-sizes, the diffusion strategies are able to deliver satisfactory performance in the presence of fading and path loss.
\end{abstract}
}
\maketitle
\IEEEdisplaynotcompsoctitleabstractindextext
\IEEEpeerreviewmaketitle

\section{Introduction}
\label{sec:intro}
\IEEEPARstart{D}{iffusion} least-mean squares (LMS) algorithms can serve as efficient and powerful mechanisms for solving distributed estimation and optimization problems over networks in real-time, in response to streaming data originating from different locations \cite{lopes2008diffusion,cattivelli2010diffusion,sayed2012diffusion,sayed2013diffusionMagazine,abdolee2013estimation}. Owing to their decentralized processing structure, simplicity of implementation, and adaptive learning capabilities, these algorithms are particularly well-suited for applications involving multi-agent wireless networks, where energy and radio resources are generally limited \cite{abdolee2011diffusion,di2011bio,abdoleedistributed}. Consensus strategies can also be used for distributed estimation purposes \cite{kar2011convergence,barbarossa2007bio,braca2008running,khan2008distributing,schizas2009distributed,aysal2009broadcast}.
However, it was shown in \cite{Tu2012outperfom} that for constant step-size adaptation, network states can grow unbounded due to an inherent asymmetry in the consensus dynamics. The same problem does not occur for diffusion strategies, and for this reason, we focus on these algorithms in this work.

Diffusion strategies have been widely investigated in networks with static topologies in which the communication links between agents remain invariant with respect to time \cite{lopes2008b,li2010,cattivelli2010diffusion,chouvardas2011adaptive,abdolee2012diffusion-process,abdoleeDiffNoisyRegreesor2012,abdolee2014diffusionNoisyRegression,
abdolee2013localization}. Under such conditions, these strategies converge in the mean and mean-square error sense in the slow adaptation regime \cite{cattivelli2010diffusion,chen2012diffusion,Tu2012outperfom,sayed2012diffusion,abdolee2013estimation}.
Previous studies have also examined the effect of noisy communication links on the performance of these algorithms on network with static topologies \cite{khalili2012transient,khalili2012steady,zhao2012imperfect,Gholami2013distributed}. The main conclusion drawn from these works is that performance degradation occurs unless the combination weights used at each node are adjusted to counter the effect of  noise.

The static link topology assumption, however, is restrictive in applications in wireless communications and sensor network systems. For example, in mobile networks where the agents are allowed to change their position over time, the signal-to-noise ratio (SNR) over the communication links between nodes will vary due to the various channel impairments, including path loss, multi-path fading and shadowing. Consequently, the set of nodes with which each agent can communicate (called neighborhood set) will also change over time, as determined by the link SNR, and the network topology is therefore intrinsically dynamic. It is therefore essential to study the performance of diffusion strategies over networks with time-varying (dynamic) topology and characterize the effects of link activity (especially link failure) on their convergence and stability.

The problem of link imperfection was also investigated in other classes of distributed algorithms, such as  consensus \cite{rabbat2005generalized,jakovetic2010weight,chan2010consensus, kar2009distributed, kar2010distributed} and subgradient algorithms  \cite{lobel2011distributed, abdoleedistributed}. In \cite{rabbat2005generalized,jakovetic2010weight} and \cite{lobel2011distributed}, the authors have examined the performance of consensus algorithms over networks with link failures, where links are established according to some predefined probabilities. They assumed that once a link is activated at a given iteration the data received through it will be undistorted.  References \cite{kar2009distributed, kar2010distributed} have taken into account the effects of link and quantization noise in addition to link failure and investigated the network convergence and stability.  A more realistic network scenario was considered in \cite{chan2010consensus,rastegarnia2014diffusion}  where the probabilities of link failure are obtained using a fading channel model and SNR of the received signals. However, the data received from a neighboring node is assumed to be error-free when the corresponding link is active.

In this paper, we study the performance of diffusion estimation strategies over networks with time-varying topologies where the information exchange between agents occurs over noisy wireless links that are also subject to fading and path loss\footnote{A short preliminary version of this work was presented in the IEEE International Conference on Communication (ICC), June 2013 \cite{abdolee2013diffusion}.}.  Our contributions are as follows. We extend the application of diffusion LMS strategies from multi-agent networks with ideal communication links to sensor networks with fading wireless channels. Under fading and path loss conditions over wireless links, the neighborhood sets become dynamic, with nodes leaving or entering  neighborhoods depending on the quality of the links as defined by the instantaneous SNR conditions. Our analysis will show that if each node knows the channel state information (CSI) of its neighbors, the effects of fading and path-loss can be mitigated by incorporating local equalization coefficients into the diffusion updates.
When CSI is not available to the nodes, we explain how the equalization coefficients can be evaluated from a pilot-assisted estimation process along with the main parameter estimation task of the network. We also examine the effect of channel estimation errors on the performance and convergence of the modified algorithms in terms of a mean-square-error metric. We establish conditions under which the network is mean-square stable for both known and unknown CSI cases. The analysis reveal that when CSI is known, the modified diffusion algorithms are asymptotically unbiased and converge  in the slow adaptation regime. In contrast, the parameter estimates will become biased when the CSI are obtained through pilot-aided channel estimation. Nevertheless, the size of the bias can be made small by increasing the number of pilot symbols or increasing the link SNR.

The paper is organized as follows. In Section \ref{sec.:networkSignalModel}, we explain the network signal model. In Section \ref{sec.:difusionLmsFadingChannel}, we review the standard diffusion strategies and introduce a modification for distributed estimation over wireless networks.
We analyze the convergence and stability of the proposed algorithms in Section \ref{sec.:DiffLmsFadingAnalysis}.
We present the simulation results in Section \ref{sec.:DiffLmsFadingResults}, and conclude the paper in Section \ref{sec.:conclusion}.

\textit{Notation:} Matrices are represented by upper-case and vectors by lower-case letters. Boldface fonts are reserved for random variables and normal fonts are used for deterministic quantities. Superscript $(\cdot)^T$ denotes transposition for real-valued vectors and matrices while $(\cdot)^{\ast}$  denotes conjugate transposition for complex-valued vectors and matrices. The symbol $\E[\cdot]$ is the expectation operator, $\text{Tr}(\cdot)$ represents the trace of its matrix argument and diag$\{\cdot\}$ extracts the diagonal entries of a matrix, or constructs a (block) diagonal matrix using its argument. A set of vectors are stacked into a column vector by $\col\{\cdot\}$. The vec$(\cdot)$ operator vectorizes a matrix by stacking its columns on top of each other and bvec($\cdot$) is the block-vectorization operator \cite{lopes2008diffusion}. The symbol $\otimes$ denotes the standard Kronecker product, and the symbol $ \otimes_{b}$ represents the \textit{block Kronecker product} \cite{lopes2008diffusion}.


\section{Network Signal Model}
\label{sec.:networkSignalModel}
Consider a set of $N$ sensor nodes that are distributed over a geographical area.
At time instant $i \in \{0,1,\cdots\}$, each node $k \in \{1,2,\cdots, N\}$ collects data $\d_k(i)$ and $\u_{k,i}$ that are related to an unknown parameter vector $w^o \in {\amsbb C}^{M\times 1}$ via the following relation:
\be
{\d_k(i)=\u_{k,i} w^o+\v_k(i)}
\label{eq.:network-data-model}
\ee
where $\d_k(i)\in {\amsbb C}$, $\u_{k,i}\in {\amsbb C}^{1 \times M}$ and $\v_k(i) \in {\amsbb C}$ are, respectively, the scalar measurement, the node's regression vector and the measurement noise.
\newtheorem{assump}{Assumption}
\begin{assump}
\label{assump.:data model}
The variables in the linear regression model (\ref{eq.:network-data-model}) satisfies the following conditions:
\begin{itemize}
  \item[a)] The regression vectors $\{\u_{k,i}\}$ are zero-mean, i.i.d. in time, and independent over space, with covariance matrices \mbox{$R_{u,k}=\E [\u_{k,i}^* \u_{k,i}]>0$}.
  \item[b)] The measurement noise $\{\v_k(i)\}$ are zero-mean, i.i.d. in time, and independent over space, with variances $\sigma^2_{v,k}$.
  \item [c)] The regression vectors $\u_{k_1,i_1}$ and the noise $\v_{k_2}(i_2)$ are mutually independent for all $k_1$, $k_2$, $i_1$ and $i_2$.
\end{itemize}
\end{assump}
\par \noindent
Node $\ell$ is said to be a neighbor of node $k$ if its distance from node $k$ is less than a preset transmission range $r_o$ \cite{see2008wireless},
which for simplicity is assumed to remain constant over the given geographical area.
The set of all neighbors of node $k$, including node $k$ itself, is denoted by ${\cal N}_k$.
Nodes are allowed to communicate with their neighbors only, but due to channel impairments,
certain links may fail.
Hence, at any given time $i$, only a subset of the nodes in ${\cal N}_k$ can
communicate with node $k$.

\begin{figure}
\centering
\includegraphics[scale=0.3]{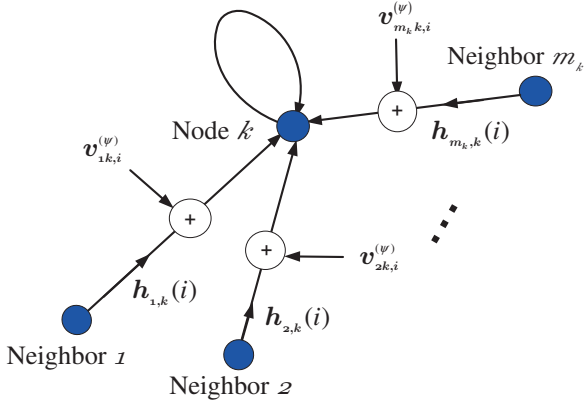}
\caption{\footnotesize{Node $k$ receives distorted data from its $m_k=|{\cal N}_{k}|$ neighbors at time $i$. The data are affected by channel fading coefficients, $\h_{\ell,k}(i)$, and communication noise $\v^{(\psi)}_{\ell k,i}$.
}}
\label{fig.:node-k-fading}
\end{figure}

The objective of the network is to estimate the unknown parameter vector $w^o$ in a distributed manner when the data exchange between the agents occurs over noisy wireless links that are also subject to fading and path loss. In particular, we assume that the transmit signal $\bpsi_{\ell,i} \in {\amsbb C}^{M\times 1}$ from node $\ell
\in {\cal N}_k\backslash \{k\}$ to node $k$ at time $i$ experiences channel distortion of the following form \mbox{(see Fig. \ref{fig.:node-k-fading})}:
\begin{align}
\bpsi_{\ell k,i}=\h_{\ell,k}(i)\sqrt{\frac{P_t}{r_{\ell,k}^{\alpha}}} \bpsi_{\ell,i}+\v_{\ell k,i}^{(\psi)}
\label{eq.:receivedWeight}
\end{align}
where $\bpsi_{\ell k,i} \in {\amsbb C}^{M\times 1}$ is the distorted estimate received by node $k$, $\h_{\ell,k}(i) \in {\amsbb C}$ denotes the fading channel coefficient over the wireless link between nodes $k$ and $\ell$, $P_t \in {\amsbb R}^+$ is the transmit signal power, $r_{\ell,k}=r_{k,\ell} \in {\amsbb R}^+$ is the distance between nodes $\ell$ and $k$, $\alpha \in {\amsbb R}^+$ is the path loss exponent and $\v_{\ell k,i}^{(\psi)} \in {\amsbb C}^{M\times 1} $ is the additive noise vector with covariance matrix  $\sigma_{v,\ell k}^{2 (\psi)}\,I_M$. We define ${\bpsi}_{k k,i} \triangleq {\bpsi}_{k,i}$ to maintain consistency in the notation.
%
%
\begin{assump}
\label{assump.:comunication_channel model}
The fading channel coefficients and the link noise in (\ref{eq.:receivedWeight}) satisfy the following conditions:
\begin{itemize}
  \item[a)] The time-varying channel coefficients $\h_{\ell,k}(i)$ follow the Clark's model \cite{proakis2000digital}, i.e., they are independent circular Gaussian random variables with zero mean and variance $\sigma^2_{h,\ell k}$.
\item[b)]  $\{\h_{\ell,k}(i)\}$ are independent over space and i.i.d. over time.
\item[c)]  The noise vectors $\{\v_{\ell k,i}^{(\psi)}\}$ are zero-mean, i.i.d. in time and independent over space. 
\item[d)] The channel coefficients, $\h_{\ell,k_1}(i_1)$, the noise vectors, $\v_{\ell k_2,i_2}^{(\psi)}$, the regression vectors, $\u_{k_3,i_3}$ and the measurement noise, $\v_{k_4}(i_4)$, are mutually independent for all $k_j$ and $i_j$ with $j \in \{ 1,2,3,4\}$.
\end{itemize}
\end{assump}
It is also assumed that nodes are aware of the positions of their neighbors through some positioning techniques and, therefore, $r_{\ell,k}$, $\ell \in {\cal N}_k$ is known to node $k$.  A transmission from node $\ell$ to node $k$ at time $i$ is said to be successful if the SNR between nodes $\ell$ and $k$, denoted by $\boldsymbol{\varsigma}_{\ell k}(i)$, exceeds some threshold level $\varsigma_{\ell k}^o$. The threshold level is defined as the SNR in the non-fading link scenario and is computed as:
\begin{align}
\varsigma^o_{\ell k} \triangleq {\frac{P_t}{\sigma^{2(\psi)}_{v,\ell k} r_{o}^{\alpha}}}
\label{eq.:snr-no-fading}
\end{align}
In fading conditions, the instantaneous SNR is:
\begin{align}
\boldsymbol{\varsigma}_{\ell k}(i)={\frac{|
\h_{\ell,k}(i)|^2P_t}{\sigma^{2(\psi)}_{v,\ell k}r_{\ell,k}^{\alpha}}}
\label{eq.:chan-threshold}
\end{align}
When transmission is successful, we have $\boldsymbol{\varsigma}_{\ell k}(i)\geq \varsigma^o_{\ell k}$ which amounts to the condition:
\begin{align}
|\h_{\ell,k}(i)|^2\geq \nu_{\ell,k}
\label{eq.succesful-reception}
\end{align}
where $ \nu_{\ell,k}=(\frac{r_{\ell,k}}{ r_{o}})^{\alpha}$. Since $\h_{\ell,k}(i)$ has a circular complex Gaussian distribution, the squared magnitude $|\h_{\ell,k}(i)|^2$ is exponentially distributed with parameter $\lambda'_{\ell, k}=1/{\sigma^2_{h,\ell k}}$ \cite{Garcia1994probability}. Considering this fact, the probability of successful transmission is then given by:
\begin{align}
p_{\ell,k}&=\textrm{Pr}\Big(|\h_{\ell,k}(i)|^2\geq \nu_{\ell,k}\Big)
=e^{-\lambda'_{\ell,k} \nu_{\ell,k}}
\label{eq.:probability-of-sucess}
\end{align}
This expression shows that the probability of successful transmission decreases as the distance between two nodes increases. As such, the link between neighboring nodes is not guaranteed to be connected all the time, implying  that the network topology  is time-varying. Under this condition, we redefine the neighborhood set of node $k$ as a time-varying set consisting of all nodes $\ell \in {\cal N}_k$ for which  $\boldsymbol{\varsigma}_{\ell k}(i)$ exceeds $\varsigma^o_{\ell k}$ provided that node $k$ knows the CSI of nodes $\ell \in {\cal N}_k$. In this way, the effective neighborhood set of each node $k$ becomes random and we, therefore, denote it by $\boldsymbol{\cal N}_{k,i}$. This implies that $\boldsymbol{\cal N}_{k,i} \subset{\cal N}_{k}$ for all $i$.

\section{Distributed Estimation over Wireless Channels}
\label{sec.:difusionLmsFadingChannel}
We first briefly review the standard diffusion LMS strategies for estimation of $w^o$ over  multi-agent networks with ideal links.
We then elaborate on how to modify these strategies to enable the estimation of $w^o$ in the presence of fading and wireless channel impairments.
\subsection{Diffusion Strategies over Ideal Communication Channels}
In the context of mean-square-error estimation, diffusion strategies are stochastic gradient algorithms that can be used for the distributed minimization of the following global objective function \cite{cattivelli2010diffusion, sayed2012diffusion}:
\begin{align}
J^{\textrm{glob}}(w)=\sum_{k=1}^N \E|\d_k(i)-\u_{k,i} w|^2
\label{eq.:globalObjFunction}
\end{align}
There are various forms of diffusion depending on the order in which the relevant adaptation and combination steps are performed. The so-called Adapt-then-Combine (ATC) strategy takes the following form:
\begin{align}
&\bpsi_{k,i}=\w_{k,i-1}+\mu_k\displaystyle \u_{k,i}^{\ast} \big [\d_{k}(i)-\u_{k,i} \w_{k,i-1}\big]\label{alg.:atc-idealLinkStep1}\\
&\w_{k,i}=\sum_{\ell \in \mathcal{N}_k} a_{\ell,k} {\bpsi}_{\ell,i}
\label{alg.:atc-idealLinkStep2}
\end{align}
%
\par \noindent
where $\mu_k>0$ is the step-size used by node $k$, and the $ a_{\ell,k}$ denote nonnegative entries of a  left-stochastic matrix $A$ that satisfy:
\begin{equation}
a_{\ell,k}=0 \;\; {\rm if} \;\; \ell \notin {\cal N}_k \;\;\;\mbox{\rm and}\;\;\;\;
\sum_{\ell\in{\cal N}_k} a_{\ell,k}=1
\label{eq.:A-conditions}
\end{equation}
In this implementation, (\ref{alg.:atc-idealLinkStep1}) is an adaptation step where node $k$ updates its intermediate estimate $\w_{k,i-1}$ to $\bpsi_{k,i}$ using its measured data $\{\u_{k,i}, \d_k(i)\}$. Then (\ref{alg.:atc-idealLinkStep2}) is a combination step in which each node $k$ combines its intermediate estimate $\bpsi_{k,i}$ with that of its neighbors to obtain $\w_{k,i}$.

While the above algorithm works well over ideal communication channels, some degradation occurs when the exchange of information between neighboring nodes is subject to noise, as explained in \cite{khalili2012transient,abdolee2011diffusion,Mateos2009consensus,Kar2009linkfailure,khalili2012steady,zhao2012imperfect}.
In this work, we move beyond these earlier studies and examine the performance of diffusion strategies over fading wireless channels. We also suggest modifications to the update equations to counter the effect of fading.

\subsection{Diffusion Strategies over Wireless Channels}
We are initially motivated to replace the combination step in (\ref{alg.:atc-idealLinkStep2}) by
\be
\w_{k,i}= \displaystyle \sum_{\ell \in {\cal N}_{k}} a_{\ell,k} {\bar \bpsi}_{\ell,i}
\label{alg.:atc-fadingLinkStep2}
\ee
where ${\bar \bpsi}_{\ell,i}$ is a refined version of the distorted estimate ${\bpsi}_{\ell k,i}$ that node $k$ receives. The refinement is computed through a scaling equalization step of the form:
\be
{\bar \bpsi}_{\ell,i}=\g_{\ell,k}(i) {\bpsi}_{\ell k,i}
\label{eq:refined-receive-data}
\ee
where the scalar gain $\g_{\ell, k}(i)$ is an equalization coefficient to be chosen to counter the effect of fading. Recall that ${\bpsi}_{\ell k,i}$ is related to $\bpsi_{\ell,i}$ via (\ref{eq.:receivedWeight}).  Moreover, since each node $k$ uses data from nodes $\ell \in {\cal N}_k$ whose instantaneous SNR, $\boldsymbol{\varsigma}_{\ell k}(i)$, exceeds the threshold ${\varsigma}_{\ell k}^o$, then we need to further adjust (\ref{alg.:atc-idealLinkStep2}) and replace ${\cal N}_k$ and $a_{\ell,k}$, respectively, with $\boldsymbol{\cal N}_{k,i}$ and $\a_{\ell,k}(i)$. This leads to:
\be
\w_{k,i}= \displaystyle \sum_{\ell \in \boldsymbol{\cal N}_{k,i}} \a_{\ell,k}(i) \g_{\ell,k}(i) {\bpsi}_{\ell k,i}
\label{eq:modified-combine-step}
\ee
Therefore, in wireless sensor networks, the ATC diffusion strategy takes the form presented in
Algorithm \ref{alg.:Diffusion-ATC-Wireless-Links}.
\begin{algorithm}
\caption{:\, Diffusion ATC over Wireless Channels}
\label{alg.:Diffusion-ATC-Wireless-Links}
\hrule
\begin{align}
&\bpsi_{k,i}=\w_{k,i-1}+\mu_k \u_{k,i}^{\ast} \big[\d_{k}(i)-\u_{k,i} \w_{k,i-1}\big] \label{alg.:atc-fadingLinkStep1}\\
&\w_{k,i}= \displaystyle \sum_{\ell \in \boldsymbol{\cal N}_{k,i}} \a_{\ell,k}(i) \g_{\ell,k}(i) {\bpsi}_{\ell k,i}
\label{alg.:atc-fadingLinkStep2}
\end{align}
\end{algorithm}
\par \noindent
One way to compute the equalization coefficients in (\ref{alg.:atc-fadingLinkStep2}) is to employ the following zero-forcing type construction:

\be
\g_{\ell, k}(i)= \left\{
  \begin{array}{ll}
   \frac{\h_{\ell,k}^*(i)}{|\h_{\ell,k}(i)|^2} \sqrt{\frac{r_{\ell,k}^{\alpha}}{P_t}}& \text{if}\; \ell \in \boldsymbol{\cal N}_{k,i}\backslash \{k\}\\
    1 & \text{if}\; \ell =k
  \end{array}
\right.
\label{eq.equalization-entries}
\ee
Alternatively, if the noise variances $\sigma_{v,\ell k}^{2(\psi)}$ are known, then one could also use minimum mean-square-error (MMSE) estimation to obtain the equalization coefficients. For simplicity, we continue with (\ref{eq.equalization-entries}). By switching the order of the adaption and combination steps in Algorithm \ref{alg.:Diffusion-ATC-Wireless-Links}, we will obtain the Combine-then-Adapt (CTA) diffusion strategy, which is presented below as Algorithm \ref{alg.:Diffusion-CTA-Wireless-Links}. In (\ref{alg.:cta-fadingLinkStep1}), $\w_{\ell k,i}$ is the estimate of the global parameter at node $\ell$ that undergoes similar path loss, fading and noise as $\bpsi_{\ell k,i}$ described by (\ref{eq.:receivedWeight}).
\begin{algorithm}
\caption{:\, Diffusion CTA over Wireless Channels}
\label{alg.:Diffusion-CTA-Wireless-Links}
\hrule
\begin{align}
&\bpsi_{k,i-1}=\displaystyle \sum_{\ell \in \boldsymbol{\cal N}_{k,i}} \a_{\ell,k}(i) \g_{\ell,k}(i) {\w}_{\ell k,i-1} \label{alg.:cta-fadingLinkStep1} \\
&\w_{k,i}=\bpsi_{k,i-1}+\mu_k \u_{k,i}^{\ast} \big[\d_{k}(i)-\u_{k,i} \bpsi_{k,i-1}\big] \label{alg.:cta-fadingLinkStep2}
\end{align}
\end{algorithm}
\par \noindent
The combination coefficients $\a_{\ell,k}(i)$ in (\ref{eq:modified-combine-step}) now become random and time-dependent because the neighborhood sets, $\boldsymbol{\cal N}_{k,i}$, are also evolving with time. Moreover, they need to satisfy
\begin{equation}
\a_{\ell,k}(i)=0 \;\; {\rm if} \;\; \ell \notin \boldsymbol{\cal N}_{k,i} \;\;\;\mbox{\rm and}\;\;\;\;
\sum_{\ell\in\boldsymbol{\cal N}_{k,i}} \a_{\ell,k}(i)=1
\label{eq.:Ai-conditions}
\end{equation}
The randomness of $\a_{\ell,k}(i)$ can be further clarified by resorting to (\ref{eq.succesful-reception}). The communication between nodes $\ell$ and $k$ is successful if (\ref{eq.succesful-reception}) is satisfied; otherwise, the link between them fails. When the link fails, the associated combination weight $\a_{\ell, k}(i)$ must be set to zero, which in turn implies that other combination coefficients of node $k$ need to be adjusted to satisfy (\ref{eq.:Ai-conditions}). This suggests that the neighborhood set $\boldsymbol{\cal N}_{k,i}$ has to be updated whenever one of the neighborhood link SNR crosses the threshold in either direction:
\begin{equation}
\boldsymbol{\cal N}_{k,i} =\Big \{\ell \in {\cal N}_k \big| \;\boldsymbol{\varsigma}_{\ell k}(i) \geq \varsigma^o_{\ell k} \Big \}
\label{eq.N-k-i}
\end{equation}
In practice, since $\boldsymbol{\varsigma}_{\ell k}(i)$ may not be measurable, we use (\ref{eq.:snr-no-fading})-(\ref{eq.:chan-threshold}) and (\ref{eq.succesful-reception}) to update the neighborhood set as:
\begin{equation}
\boldsymbol{\cal N}_{k,i} =\Big \{\ell \in {\cal N}_k \big| \;|\h_{\ell k}(i)|^2 \geq \nu_{\ell k} \Big \}
\label{eq.N-k-i-v2}
\end{equation}
Motivated by these considerations, we propose the following  dynamic structure to adjust the combination weights over time:
\be
\a_{\ell,k}(i)
=\left\{
\begin{array}{ll}
\gamma_{\ell,k} \boldsymbol{\cal I}_{\ell,k}(i),&\quad {\rm if} \; \ell  \in \boldsymbol{\cal N}_{k,i}\backslash \{k\} \\
1-\sum_{\ell \in \boldsymbol{\cal N}_{k,i} \backslash\{k\}} \a_{\ell,k}(i),&\quad {\rm if} \; \ell=k\\
\end{array}
\label{eq.:Ai}
\right.
\ee
where the $\gamma_{\ell,k}$ are fixed, positive combination weights that node $k$ assigns to its neighbors $\ell \in \boldsymbol{\cal N}_{k,i}$. To ensure $\a_{k,k}(i)>0$, these weights need to satisfy:
\be
\sum_{\ell \in \boldsymbol{\cal N}_{k,i}\backslash \{k\}} \gamma_{\ell,k} < 1
\label{eq.:AfixCondtion}
\ee
It can be verified that if each node $k$ obtains the coefficients $\gamma_{\ell,k}$ for the time-invariant neighborhood set ${\cal N}_k$   according to well-known left or doubly-stochastic  matrix combination rules (e.g., uniform averaging rule or Metropolis rule) then the condition (\ref{eq.:AfixCondtion}) will be satisfied.
In (\ref{eq.:Ai}), the quantity  $\boldsymbol{\cal I}_{\ell,k}(i)$ is defined as:
\vspace{-2mm}
\begin{align}
\boldsymbol{\cal I}_{\ell,k}(i)=
\left\{\begin{array}{l}
1,\quad {\rm if} \; \ell  \in \boldsymbol{\cal N}_{k,i}\\
0, \quad \; {\rm otherwise}
\end{array}
\label{eq.:Ical-diffusionLms}
\right.
\end{align}
When transmission from node $\ell$ to node $k$ is successful $\boldsymbol{\cal I}_{\ell,k}(i)=1$, otherwise, $\boldsymbol{\cal I}_{\ell,k}(i)=0$. In this way, the entries $\a_{\ell,k}(i)$ satisfy condition (\ref{eq.:Ai-conditions}). From (\ref{eq.N-k-i}) and (\ref{eq.:Ical-diffusionLms}), we see that the indicator operator, $\boldsymbol{\cal I}_{\ell,k}(i)$, is a random variable with Bernoulli distribution for which the probability of success, $p_{\ell,k}$, is given by the exponential function (\ref{eq.:probability-of-sucess}).
\subsection{Modeling the Impact of Channel Estimation Errors}
\label{subsec.channel-estimstion}
In Algorithms \ref{alg.:Diffusion-ATC-Wireless-Links} and \ref{alg.:Diffusion-CTA-Wireless-Links},
it is assumed that each node $k$ knows the channel fading coefficients $\h_{\ell,k}(i)$, which are needed in (\ref{eq.equalization-entries}). In practice, this information is usually recovered by means of an estimation step. Consequently, some additional estimation errors will be introduced into the network.

There are many ways by which the fading coefficients can be estimated. For example, we may assume that the transmitted data from node $\ell$ to node $k$ carries two data types, namely, pilot symbols (training data) denoted by $\s_{\ell}(i)$, and  data symbols $\bpsi_{\ell,i}$ or $\w_{\ell,i-1}$. The training data are used for channel estimation and the data symbols are the intermediate estimates of the unknown parameter vector, $w^o$, which are used  to update the network estimate at node $k$.
According to (\ref{eq.:receivedWeight}), the received training data at node $k$ and time $i$ is affected by fading and noise, i.e.,
\begin{align}
\y_{\ell, k}(i)=\h_{\ell,k}(i)\sqrt{\frac{P_t}{r_{\ell,k}^{\alpha}}} \s_{\ell}(i)+\v_{\ell, k}^{(y)}(i)
\label{eq.:ReceivedPilotData}
\end{align}
where $\v_{\ell, k}^{(y)}(i)$ is a zero-mean additive white Gaussian noise with variance $\sigma_{v,\ell k}^{(y)2}$. It is reasonable to assume that $\sigma_{v,\ell k}^{(y)2}=\sigma_{v,\ell k}^{(\psi)2}$. The number of training symbols used depends on the specific application requirements and the time scale variations of the channel.
If we use a single training data to estimate each coefficient and assume that nodes $k\in \{1,2,\cdots,N\}$ sends $\s_{k}(i)=1$  as training symbols, the least-squares estimation method gives the following estimate:
\be
\hat{\h}_{\ell,k}(i)=\sqrt{\frac{r_{\ell,k}^{\alpha}}{P_t}}\y_{\ell,k}(i)
\label{eq.:estimatedChannel-v2}
\ee

\newtheorem{remark}{Remark}
\begin{remark}
\label{re:remark--1}
If we use an alternative way to find the \mbox{threshold} SNR, $\nu^o_{\ell,k}$  without using distance information, then (\ref{eq.:ReceivedPilotData}) can be expressed as $ \y_{\ell, k}(i)=\boldsymbol{\beta}_{\ell,k} (i)
\s_{\ell}(i)+\v_{\ell, k}^{(y)}(i)$, where $\boldsymbol{\beta}_{\ell,k}(i)=\h_{\ell,k}(i) (P_t/ r^{\alpha}_{\ell,k})^{1/2}$.  In this form the fading coefficient and path loss are combined into a new channel coefficient $\boldsymbol{\beta}_{\ell,k} (i)$  that implicitly includes the distance information. In this case, to estimate the channel coefficients, $\boldsymbol{\beta}_{\ell,k} (i)$, unlike  (\ref{eq.:estimatedChannel-v2}), the distance information are not required.
\end{remark}

From (\ref{eq.:ReceivedPilotData}), it can be seen that $\y_{\ell, k}(i)$ is composed of the sum of two independent circular Gaussian random variables.  It follows that $\y_{\ell, k}(i)$ will have circular Gaussian distribution with zero mean and variance $\sigma^2_{h,\ell k}\frac{P_t}{r_{\ell,k}^{\alpha}}+\sigma_{v,\ell k}^{(\psi)2}$. From (\ref{eq.:estimatedChannel-v2}), we therefore conclude that $\hat{\h}_{\ell,k}(i)$ has circular Gaussian distribution with zero mean and variance  $\sigma^2_{h,\ell k}+\frac{r_{\ell,k}^{\alpha}}{P_t} \sigma_{v,\ell k}^{(\psi)2}$, and $|\hat{\h}_{\ell,k}(i)|^2$ has exponential distribution with parameter
\be
\lambda_{\ell,k}=\,\frac{1}{\sigma^2_{h,\ell k}+\frac{r_{\ell,k}^{\alpha}}{P_t}\sigma_{v,\ell k}^{(\psi)2}}
\label{eq.:LambdaForh}
\ee
From here the probability of successful transmission from node $\ell$ to node $k$ will be defined in terms of the estimated channel coefficient as
\begin{align}
p_{\ell,k}& \triangleq \textrm{Pr}\Big(| \hat{\h}_{\ell,k}(i)|^2\geq  \nu_{\ell,k}\Big)
=e^{-\lambda_{\ell,k}\, \nu_{\ell,k}}
\label{eq.:probability-of-sucess-imperfect}
\end{align}
Considering the assumed training data and from (\ref{eq.:ReceivedPilotData}) and (\ref{eq.:estimatedChannel-v2}), the instantaneous channel estimation error will be
\begin{align}
{\tilde \h}_{\ell,k}(i)=\h_{\ell,k}(i)-{\hat \h}_{\ell,k}(i)
=\;-\sqrt{\frac{r_{\ell,k}^{\alpha}}{P_t}} \v_{\ell, k}^{(y)}(i)
\end{align}
Therefore, the variance of the estimation error is:
\be
\sigma^2_{{\tilde h}_{\ell,k}}=\E|{\tilde \h}_{\ell,k}(i)|^2= \frac{r_{\ell,k}^{\alpha}}{P_t} \sigma^{(\psi) 2}_{v,\ell k}
\ee
which shows that the power of the channel estimation error, $\sigma^2_{{\tilde h}_{\ell,k}}$, decreases if the node transmit power increases or if the distance between nodes $\ell$ and $k$ decreases. To reduce the channel estimation error, the alternative solution is to use more pilot data. It can be shown that if the wireless channel remains invariant over the transmission of $n$  pilot data, then the estimation error variance will be scaled by a factor of $1/n$ \cite{kay1993fundamentals}.

\begin{remark}
\label{re:remark-0}
The time index $i$, in Algorithms  \ref{alg.:Diffusion-ATC-Wireless-Links} and \ref{alg.:Diffusion-CTA-Wireless-Links}, refers to the iteration number of adaptation and combination steps and not the time at which the communication between nodes occurs. This implies that from time index $i-1$ to $i$, a node may transmit several training symbols to its neighbors for channel estimation process and, therefore, the estimated channels used in iteration $i$ may be obtained using several pilot data. However, to simplify the presentation, we  also use index $i$ to represent the communication time of pilots in (\ref {eq.:ReceivedPilotData}) since it is assumed that a single pilot datum used for channel estimation.
\end{remark}

We can now express  (\ref{eq.:receivedWeight}) in terms of the estimated channels ${\hat \h}_{\ell,k}(i)$ and the channel estimation error as
\begin{align}
\bpsi_{\ell k,i}={\hat \h}_{\ell,k}(i) \sqrt{\frac{P_t} {r_{\ell,k}^{\alpha}}}  \bpsi_{\ell,i}
+\sqrt{\frac{P_t} {r_{\ell,k}^{\alpha}}} {\tilde \h}_{\ell,k}(i) \bpsi_{\ell,i}+\v_{\ell k,i}^{(\psi)}
\label{eq.:receivedWeightVer2}
\end{align}
The equalization coefficients ${\hat \g}_{\ell,k}(i)$ are computed using the estimated channels ${\hat \h}_{\ell,k}(i)$, according to (\ref{eq.equalization-entries}). Using this construction, the equalized received data at node $k$ become:
\begin{align}
{\hat \g}_{\ell, k}(i) \bpsi_{\ell k,i}
           &=\Big(1+{\hat \g}_{\ell,k}(i) \sqrt{\frac{P_t} {r_{\ell,k}^{\alpha}}} {\tilde \h}_{\ell k}(i)\Big) \bpsi_{\ell,i}\nonumber \\
           &\hspace{2cm}+{\hat \g}_{\ell,k}(i) \v_{\ell k,i}^{(\psi)}
\label{eq.:equalizedWeight}
\end{align}
Substituting the equalized data into (\ref{alg.:atc-fadingLinkStep2}), we obtain:
\be
\w_{k,i}=\sum_{\ell\in \boldsymbol{\cal N}_{k,i}} \a_{\ell,k}(i) \bpsi_{\ell,i}+ \sum_{\ell\in \boldsymbol{\cal N}_{k,i}} \e_{\ell,k}(i) \bpsi_{\ell,i}+\v_{k,i}^{(\psi)}
\label{alg.:atc-fadingLinkStep3-v2}
\ee
where
\vspace{-3mm}
\begin{align}
\e_{\ell,k}(i)&=-\a_{\ell,k}(i) {\hat \g}_{\ell,k}(i) \v^{(y)}_{\ell k}(i) \\
\v^{(\psi)}_{k,i}&=\sum_{\ell\in \boldsymbol{\cal N}_{k,i}}\a_{\ell,k}(i){\hat \g}_{\ell,k}(i) \v^{(\psi)}_{\ell k,i}
\end{align}

There are several important features in the combination step (\ref{alg.:atc-fadingLinkStep3-v2}) that need to be highlighted.
First, the combination coefficients,  $\a_{\ell,k}(i)$, used in this step are time varying. These  coefficients, in addition to combining the exchanged information, model the link failure phenomenon over the network. Second, $\{{\hat \g}_{\ell,k}(i)\}$ account for the effects of fading channels. Using these variables and the control SNR mechanism introduced above, we can reduce the effect of  link noise. Third, in (\ref{alg.:atc-fadingLinkStep3-v2}), $\{\e_{\ell k}(i)\}$  model the channel estimation errors, which allows us to examine the impact of these errors on the diffusion strategies.

In summary, in a multi-agent wireless network, each node $k$ will perform the processing tasks listed in Table I in order of precedence to complete cycle $i$ of the ATC diffusion LMS algorithm.
\begin{algorithm}
\caption*{{TABLE I:\; ATC diffusion implementation}}
\label{alg.:summary-task}
\hrule
\begin{align}
&\hat{\h}_{\ell,k}(i)=\left\{
\begin{array}{ll}
\sqrt{\frac{r_{\ell,k}^{\alpha}}{P_t}}  \, \y_{\ell,k}(i) & \text{if}\; \ell \in {\cal N}_k \backslash \{k\}\\
1 & \text{if}\; \ell =k
  \end{array}
\label{eq.ta-hat-hlk-i}
\right.
\\
&\boldsymbol{\cal N}_{k,i} =\Big \{\ell \in {\cal N}_k \big| \;\hat{\h}_{\ell, k}(i) \geq \nu_{\ell k} \Big \}
\label{eq.ta-Nki} \\
&\hat{\g}_{\ell, k}(i)= \left\{
  \begin{array}{ll}
\frac{\hat{\h}_{\ell,k}^*(i)}{|\hat{\h}_{\ell,k}(i)|^2} \sqrt{\frac{r_{\ell,k}^{\alpha}}{P_t}}& \text{if}\; \ell \in \boldsymbol{\cal N}_{k,i}\backslash \{k\}\\
    1 & \text{if}\; \ell =k\\
  \end{array}
\label{eq.ta-hat-glk-i}
\right.
\\
&\boldsymbol{\cal I}_{\ell,k}(i)=
\left\{\begin{array}{l}
1,\quad {\rm if} \; \ell  \in \boldsymbol{\cal N}_{k,i}\\
0, \quad \; {\rm otherwise}\\
\end{array}
\right. \\
&\a_{\ell,k}(i)
=\left\{
\begin{array}{l}
\gamma_{\ell,k} \boldsymbol{\cal I}_{\ell,k}(i),\quad {\rm if} \; \ell  \in \boldsymbol{\cal N}_{k,i}\backslash \{k\}\\
1-\sum_{\ell \in \boldsymbol{\cal N}_{k,i} \backslash\{k\}} \a_{\ell,k}(i),\quad {\rm if} \; \ell=k
\end{array}
\right.\\
&\bpsi_{k,i}=\w_{k,i-1}+\mu_k \u_{k,i}^{\ast} \big[\d_{k}(i)-\u_{k,i} \w_{k,i-1}\big] \label{alg.:atc-fadingLinkStep1}\\
&\w_{k,i}= \displaystyle \sum_{\ell \in \boldsymbol{\cal N}_{k,i}} \a_{\ell,k}(i) \g_{\ell,k}(i) {\bpsi}_{\ell k,i}
\label{alg.:atc-fadingLinkStep2}
\end{align}
\end{algorithm}

\section{Performance Analysis}
\label{sec.:DiffLmsFadingAnalysis}
In this section, we derive conditions under which the equalized diffusion strategies are stable in the mean and mean square sense.
We also derive expressions to characterize the  mean-square-deviation (MSD)  and excess mean-square-error (EMSE) performance levels  of the algorithms during the transient phase and in steady-state. We focus on the ATC variant (\ref{alg.:atc-fadingLinkStep1})--(\ref{alg.:atc-fadingLinkStep2}). The same conclusions hold for
 (\ref{alg.:cta-fadingLinkStep1})-(\ref{alg.:cta-fadingLinkStep2}) with minor adjustments.

To derive a recursion for the mean error-vector of the network, we begin with defining the local error vectors:
\begin{align}
\tilde \w_{k,i}\triangleq w^o-\w_{k,i} \\
\tilde{\bpsi}_{k,i}\triangleq w^o-\bpsi_{k,i}
\end{align}
We subtract $w^o$ from both sides of (\ref{alg.:atc-fadingLinkStep1}) and (\ref{alg.:atc-fadingLinkStep3-v2}) to obtain:
\begin{align}
&{\tilde \bpsi}_{k,i}=(I-\mu_k \u_{k,i}^{\ast} \u_{k,i}) {\tilde \w}_{k,i-1}-\mu_k \u_{k,i}^{\ast} \v_{k}(i) \label{alg.:atc-errorRecursion1} \\
&{\tilde \w}_{k,i}=\sum_{\ell \in \boldsymbol{\cal N}_{k,i}} \a_{\ell,k}(i) {\tilde \bpsi}_{\ell,i}+ \sum_{\ell \in \boldsymbol{\cal N}_{k,i}} \e_{\ell,k}(i) {\tilde \bpsi}_{\ell,i} \nonumber \\
&\hspace{2cm} +\sum_{\ell \in \boldsymbol{\cal N}_{k,i}} \e_{\ell,k}(i) w^o-\v_{k,i}^{(\psi)}
\label{alg.:atc-errorRecursion2}
\end{align}
We collect the $\{\a_{\ell,k}(i)\}$ into a left-stochastic matrix $\A_i$ and the $\{\e_{\ell,k}(i)\}$ into an error matrix $\boldsymbol{E}_i$. We also define  the extended versions of these matrices using Krocecker products as  $\boldsymbol{\mathcal A}_i\triangleq \boldsymbol{A}_i\otimes I_{M}$ and $\boldsymbol{\mathcal E}_i \triangleq \boldsymbol{E}_i \otimes I_{M}$.
We further introduce the network error vectors:
\begin{align}
&\tilde{\bpsi}_i\triangleq\col\{\tilde{\bpsi}_{1,i},\tilde{\bpsi}_{2,i},\ldots,\tilde{\bpsi}_{N,i}\}\\
&\tilde \w_i\triangleq\col\{\tilde\w_{1,i},\tilde\w_{2,i},\ldots,\tilde \w_{N,i}\}
\label{eq:tilde w_def}
\end{align}
and the variables:
\begin{align}
&\boldsymbol{\mathcal{R}}_i\triangleq \text{diag}\Big \{\u_{1,i}^{\ast}{\u}_{1,i},\cdots,\u_{N,i}^{\ast}{\u}_{N,i} \Big\} \\
&{\cal M}\triangleq\text{diag}\Big \{ \mu_1 I_{M}, \cdots, \mu_N I_{M} \Big \} \\
&\boldsymbol{p}_i\triangleq\text{col}\big\{{\u}_{1,i}^{\ast}\v_1(i),\cdots,{\u}_{N,i}^{\ast}\v_N(i)\big\} \label{eq.:Gi-definition}\\
&\v^{(\psi)}_i\triangleq\text{col}\big\{\v^{(\psi)}_{1,i},\cdots,\v^{(\psi)}_{N,i}\big\} \label{eq.:vi-definition} \\
&\omega^o \triangleq {\mathbb 1}_N \otimes w^o \label{eq.:omegao-definition}
\end{align}
where ${\mathbb 1}_N$ is a column vector with length $N$ and unit entries. We can now use (\ref{alg.:atc-errorRecursion1}) and (\ref{alg.:atc-errorRecursion2}) to verify that the following recursion holds for the network error vector:
\begin{align}
\tilde\w_i= \boldsymbol{\cal{B}}_i\tilde\w_{i-1}-({\boldsymbol{\mathcal A}}_i+{\boldsymbol{\mathcal E}_i})^T{\cal M} \boldsymbol{{p}}_i+{\boldsymbol{\mathcal E}_i}^T\omega^o-\v^{(\psi)}_i
\label{eq.:global_error_vector_w}
\end{align}
where
\be
\boldsymbol{{\cal B}}_i=({\boldsymbol{\mathcal A}}_i+{\boldsymbol{\mathcal E}_i})^T(I-{\cal M} \boldsymbol{\mathcal{R}}_i)
\label{eq:calBi}
\ee
\subsection{Mean Convergence}
Taking the expectation of (\ref{eq.:global_error_vector_w}) under Assumptions \ref{assump.:data model} and \ref{assump.:comunication_channel model}, we arrive at
\begin{align}
\E[\tilde\w_i]={\cal B}\, \E [\tilde\w_{i-1}]+{{\mathcal E}}^T\omega^o
\label{eq.:mean_perfomance}
\end{align}
where
\begin{align}
&{\cal B}\triangleq \E[\boldsymbol{\cal B}_i]=({\mathcal A}+{\cal E})^T(I-{\cal M}{\mathcal R})\\
&{\cal A} \triangleq \E[\boldsymbol{\cal A}_i]=A \otimes I_M\\
&{\cal E} \triangleq \E[\boldsymbol{\cal E}_i]=E \otimes I_M\\
&{\mathcal R}\triangleq \E[\boldsymbol{\cal R}_i]=\diag \big \{ R_{u,1} ,\ldots,  \, R_{u,N} \big \}
\end{align}
To obtain (\ref{eq.:mean_perfomance}), we used the fact that $ \v_k(i)$ is independent of $\u_{k,i}$ and $\E[\v_k(i)]=0$. Moreover, we have
$\E[\v^{(\psi)}_i]=0$ because  $\hat{\g}_{\ell,k}(i)$ is independent of $\v^{(\psi)}_{\ell k,i}$ and $\E[\v^{(\psi)}_{\ell k,i}]=0$. Considering the time-varying left-stochastic matrix $\A_i$, we can use (\ref{eq.:Ai}) to find the entries of
$A=\E[\A_i]$, i.e.,
\be
a_{\ell,k}
=\left\{
\begin{array}{ll}
\gamma_{\ell,k} p_{\ell,k}, &\quad {\rm if} \; \ell  \in {\cal N}_k \backslash \{k\} \\
1-\sum_{ \ell \in {\cal N}_k \backslash \{k\} } \gamma_{\ell,k} p_{\ell,k},&\quad {\rm if} \; \ell=k
\end{array}
\right.
\ee
Observe that  $A^T {\mathbb 1}={\mathbb 1}$.
The $(\ell,k)$-th entry of matrix $E$ is zero on the diagonal and, for $\ell\neq k$, is given by:
\begin{align}
e_{\ell,k}&=-\E\big[\a_{\ell,k}(i) {\hat \g}_{\ell,k}(i) \v_{\ell,k}^{(y)}(i)\big] \nonumber \\
&\numrel{=}{label-1}-\gamma_{\ell,k}\,\E \big[\boldsymbol{\cal I}_{\ell,k}(i){\hat \g}_{\ell,k}(i) \v_{\ell,k}^{(y)}(i)\big]
\nonumber \\
&\numrel{=}{label-2}-\gamma_{\ell,k}\, \E\Big[{\hat \g}_{\ell,k}(i) \v_{\ell,k}^{(y)}(i)\Big | |\hat{\h}_{\ell, k}(i)|^2 \geq  \nu_{\ell,k}\Big] \nonumber \\
&\numrel{=}{label-3}-\gamma_{\ell,k}\, \,\E \Bigg[\Big(\frac{\sqrt{\frac{r^{\alpha}}{P_t}}{\h}^*_{\ell,k}(i) \v_{\ell,k}^{(y)}(i)+\frac{r^{\alpha}}{P_t}|\v_{\ell,k}^{(y)}(i)|^2}{|{\h}_{\ell,k}(i)
+\sqrt{\frac{r^{\alpha}}{P_t}}\v_{\ell,k}^{(y)}(i)|^2}\Big) \nonumber\\
& \hspace{1.5cm} \Bigg | \Big(\big|{\h}_{\ell,k}(i)+\sqrt{\frac{r^{\alpha}}{P_t}}\v_{\ell,k}^{(y)}(i)\big|^2 \geq  \nu_{\ell,k}\Big)\Bigg]
\label{eq.:expectationOfE-4}
\end{align}
The equality in step (\ref{label-2}) follows from the fact that ${\hat \g}_{\ell,k}(i)$ is defined for $\ell \in {\cal N}_{k}\backslash \{k\}$ when $|\hat{\h}_{\ell, k}(i)|^2 \geq  \nu_{\ell,k}$, for which $\Ical_{\ell,k} (i)=1$. We obtain (\ref{label-3}) by expressing ${\hat \g}_{\ell,k}(i)$ in terms of ${\h}_{\ell,k}(i)$ and $\v_{\ell,k}^{(y)}(i)$ according to (\ref{eq.:ReceivedPilotData}), (\ref{eq.ta-hat-hlk-i}) and (\ref{eq.ta-hat-glk-i}). Expression (\ref{eq.:expectationOfE-4}) indicates that $e_{\ell,k}$ is bounded.

\begin{remark}
\label{re:remark-1}
From the right hand side of (\ref{eq.:expectationOfE-4}),   it can be verified that the value of the expectation is independent of time since the estimation error, $\v_{\ell,k}^{(y)}(i)$, and the channel coefficients, ${\h}_{\ell,k}(i)$, are assumed to be i.i.d. over time with fixed probability density functions.
\end{remark}
According to (\ref{eq.:mean_perfomance}), when ${\cal B}$ is stable, then the network mean error vector converges to
\begin{align}
b\triangleq\lim_{i\rightarrow \infty} \E[\tilde \w_i]=(I-{\cal B})^{-1} {\mathcal E}^T\omega^o
\label{eq.:MeanSteadyState}
\end{align}
If $\hat \h_{\ell, k}(i)=\h_{\ell, k}(i)$ then ${\mathcal E}=0$ and $\lim_{i\rightarrow \infty} \E[\tilde \w_i]=0$, i.e., the algorithm will be asymptotically unbiased.

Let us now find conditions under which ${\cal B}$ is stable, i.e., conditions under which the spectral radius of ${\cal B}$, denoted by $\rho({\cal B})$, is strictly less than one. We use the properties of
the block maximum norm $\|\cdot\|_{b,\infty}$ from \cite{sayed2012diffusion,takahashi2010diffusion} to establish the following relations:
\begin{align}
\rho({\cal B}) &\leq \|{\cal B}\|_{b,\infty} \nonumber \\
&\leq \|({\cal A}+{\cal E})^T \|_{b,\infty}\, \|(I-{\cal M}{\cal R})\|_{b,\infty} \nonumber \\
&\leq \Big(\|{\cal A}^T\|_{b,\infty}+\|{\cal E}^T\|_{b,\infty} \Big)\, \|(I-{\cal M}{\cal R})\|_{b,\infty} \nonumber \\
&= \big (1+\|{\cal E}^T\|_{b,\infty} \big ) \|(I-{\cal M}{\cal R})\|_{b,\infty} \label{eq.:mean-stabilty-inequality4}
\end{align}
%
where in the last equality we used the fact that  $\|{\cal A}^T\|_{b,\infty}=1$ since $A$ is left-stochastic.  According to (\ref{eq.:mean-stabilty-inequality4}), $\rho({\cal B})$ is bounded by one if
\begin{align}
\|(I-{\cal M}{\cal R})\|_{b,\infty}<\frac{1}{1+\|{\cal E}\|_{b,\infty}}
\end{align}
Since $I-{\cal M}{\cal R}$ is block diagonal and Hermitian, we have $\|(I-{\cal M}{\cal R})\|_{b,\infty}=\rho(I-{\cal M}{\cal R})$ \cite{sayed2012diffusion}. The spectral radius of $I-{\cal M}{\cal R}$ will be less than ${1}/{(1+\|{\cal E}\|_{b,\infty})}$ if the absolute maximum eigenvalue of each of its blocks is strictly less than ${1}/{(1+\|{\cal E}\|_{b,\infty})}$. This condition is satisfied if at each node $k$ the step-size $\mu_k$ is chosen as:
\begin{equation}
\frac{1-\frac{1}{1+\|{\cal E}\|_{b,\infty}}}{\lambda_{\max}(R_{u,k})}<\mu_k < \frac{1+\frac{1}{1+\|{\cal E}\|_{b,\infty}}}{\lambda_{\max}(R_{u,k})}
\label{eq.:mu_range}
\end{equation}
where $\lambda_{\max}(\cdot)$ denotes the maximum eigenvalue of its matrix argument. This relation reveals that the mean-stability range of the algorithm, in terms of the step size parameters $\{\mu_k\}$, reduces as the channel estimation error over the network increases.
When the channel estimation error approaches zero\footnote{The channel estimation error can be reduced by transmitting more  pilot symbols or increasing the SNR during pilot transmission.}, that is when $\|{\cal E}\|_{b,\infty} \rightarrow 0$, the stability condition reduces to \mbox{$0<\mu_k< \frac{2}{\lambda_{\max}(R_{u,k})}$}, which is the mean stability range of diffusion LMS over ideal communication links\cite{sayed2012diffusion}.  A similar analysis can be carried out for the CTA diffusion strategy.

\newtheorem{thm}{Theorem}
\begin{thm}
\label{thm.:MeanPerformanceDLMSfading}
Consider the diffusion strategies (\ref{alg.:atc-fadingLinkStep1})--(\ref{alg.:atc-fadingLinkStep2})  with the space-time data (\ref{eq.:network-data-model}) and (\ref{eq.:receivedWeight}) satisfying Assumptions \ref{assump.:data model} and \ref{assump.:comunication_channel model}, respectively, and where the channel coefficients are estimated using (\ref{eq.:estimatedChannel-v2}) with training symbols $ s_k(i)=1$.
Then the algorithms will be stable in the mean and the mean error vector will converge to (\ref{eq.:MeanSteadyState}) if the step-sizes are  chosen according to (\ref{eq.:mu_range}).
\end{thm}

\setcounter{equation}{79}
\begin{figure*}[!b]
\vspace{5pt}
\hrulefill
\begin{align}
\bar{\cal F}=&(I-{\cal M}{\cal R})^T \otimes_b (I-{\cal M} {\cal R})+\Bigg \{\sum_{k=1}^N \Big[ \diag \big((\vec(\diag(e_k))\big) \Big]\otimes \Big[ (\beta-1)( R^T_{k,u} \otimes R_{k,u})+r_k r^*_k \Big] \Bigg \} ({\cal M} \otimes_b {\cal M})
\label{eq:bar-F-Gussian-2}
\end{align}
\end{figure*}
\setcounter{equation}{66}
\subsection{Steady-State Mean-Square Performance}
\label{subsec.:MeanSquareAnalysis}
To study the mean-square performance of the algorithm, we need to determine the network variance relation \cite{sayed2008,lopes2008diffusion,zhao2012imperfect}. The latter can be obtained by equating the weighted squared norms of both sides of (\ref{eq.:global_error_vector_w}), and taking expectations under Assumptions \ref{assump.:data model} and \ref{assump.:comunication_channel model}:
\begin{align}
&\E\|\tilde \w_i\|^2_{\Sigma}=\E \|\tilde \w_{i-1}\|^2_{{\boldsymbol{\Sigma}_i}'}+\E \big[\omega^{o*} \, \boldsymbol{\cal E}^{*T}_i \Sigma \boldsymbol{\cal E}_i^T \, \omega^o \big] \nonumber \\
&\hspace{0.1cm}+\E \big [\boldsymbol{p}^{\ast}_i {\cal M}^T (\boldsymbol{\cal A}_i+\boldsymbol{\cal E}^{*T}_i)\Sigma (\boldsymbol{\cal A}_i+\boldsymbol{\cal E}_i)^T{\cal M}\boldsymbol{p}_i \big] \nonumber \\
&\hspace{0.1cm}+2\text{Re}\Big\{\E[ \omega^{o*} \boldsymbol{\cal E}^{*T}_i \Sigma\boldsymbol{\cal B}_i {\tilde \w}_{i-1}]\Big\}+\E[\v^{(\psi)*}_i \Sigma \v^{(\psi)}_i]
\label{variance_relation_1}
\end{align}
where for a vector $x$ and a weighting matrix $\Sigma \geq0$ with compatible dimensions $\|x\|^2_{\Sigma}=x^* \Sigma x$, and
\begin{align}
{\boldsymbol{\Sigma}}'_i={\boldsymbol{{\cal B}}_i}^{\ast} \Sigma \boldsymbol{{\cal B}}_i
\end{align}
Under the independence assumption between ${\tilde \w}_{i-1}$ and $\boldsymbol{\cal R}_i$, it holds that
\begin{align}
\E\big[\|\tilde \w_{i-1}\|^2_{{\boldsymbol{\Sigma}}'_i}\big]=\E\|\tilde \w_{i-1}\|^2_{\E[{\boldsymbol{\Sigma}}'_i]}
\end{align}
Using this equality in (\ref{variance_relation_1}), we arrive at:
\begin{align}
\E&\|\tilde \w_i\|^2_{\Sigma}=\E\|\tilde \w_{i-1}\|^2_{{\Sigma}'}+\Tr(\E \big[\boldsymbol{\cal E}_i^T \omega^{o*}\, \omega^o \boldsymbol{\cal E}^{*T}_i \Sigma] \big) \nonumber \\
&+\Tr \big(\E[(\boldsymbol{\cal A}_i+\boldsymbol{\cal E}_i)^T {\cal M} \boldsymbol{p}_i \boldsymbol{p}_i^*{\cal M} (\boldsymbol{\cal A}_i+\boldsymbol{\cal E}^{*T}_i) \Sigma ]\big)\nonumber \\
&+2\text{Re}\Big\{\Tr(\E[\boldsymbol{\cal B}_i {\tilde \w}_{i-1} \omega^{o*} \boldsymbol{\cal E}^{*T}_i \Sigma])\Big\} +\Tr\big(\E[\v^{(\psi)}_i \v^{(\psi)*}_i \Sigma]\big)
\label{variance_relation_2}
\end{align}
where ${\Sigma}'=\E[{\boldsymbol{\Sigma}}'_i]$. To compute (\ref{variance_relation_2}), we introduce:
\begin{align}
&{\cal P}       \triangleq     \E[\boldsymbol{p}_i \boldsymbol{p}_i^*]=\diag \big \{\sigma^2_{v,1} R_{u,1},\cdots,\sigma^2_{v,N} R_{u,N}\big \} \\
&{\cal R}_v     \triangleq     \diag\big\{R_{v,1}\cdots, R_{v,N}\big\}\\
&R_{v,k} \triangleq  \E[\v^{(\psi)}_{k,i} \v^{(\psi)*}_{k,i}]=\sum_{\ell \in {\cal N}_{k} \backslash \{k\}} \E \big[\a^2_{\ell,k}(i) \,|{\hat \g}_{\ell,k}(i)|^2\big] R^{(\psi)}_{v,\ell k}
\label{eq:Rvlk}
\end{align}
We show in Appendix \ref{Appdix:Computation-of-Rvk} how to compute the expectation term multiplying $R_{v,\ell k}^{(\psi)}$ in (\ref{eq:Rvlk}). Alternatively, this term can be evaluated numerically by averaging over repeated independent experiments.

To proceed, we assume that $\Sigma$ is partitioned into block entries of size $M\times M$ and let $\sigma=\mbox{\rm bvec}(\Sigma)$ denote the vector that is obtained from the block vectorization of $\Sigma$. We shall write $\|\tilde{\w}_i\|^2_{\Sigma}$ and $\|\tilde{\w}_i\|_{\sigma}^2$ interchangeably to denote the same weighted square norm \cite{lopes2008diffusion}.  Using properties of bvec and block Kronecker products \cite{koning1991block}, the variance relation in (\ref{variance_relation_2}) leads in steady-state to:
\begin{align}
\lim_{i \rightarrow \infty }\E\|\tilde \w_i\|^2_{\sigma}=&\lim_{i \rightarrow \infty }\E\|\tilde \w_{i-1}\|^2_{{{\cal F}\sigma}}+\gamma^T \sigma
\label{variance_relation_3}
\end{align}
where ${\cal F} = \E[\boldsymbol{\cal B}_i^T   \otimes_{b} \boldsymbol{\cal B}^*_i]$, and
\begin{align}
&\gamma=\lim_{i \rightarrow \infty } \Big\{\E\big[\boldsymbol{\cal E}_i^T   \otimes_{b}
\boldsymbol{\cal E}_i^* \big]\,\bvec\big((\omega^{o} \omega^{o*})^T\big) \nonumber \\
&+ \E\big[(\boldsymbol{\cal A}_i+\boldsymbol{\cal E}_i)^T   \otimes_{b} (\boldsymbol{\cal A}_i+\boldsymbol{\cal E}^{*T}_i)^T\big]\, \bvec({\cal M} {\cal P}^T {\cal M}) \nonumber \\
&+2\text{Re}\{\E\big[\boldsymbol{\cal B}_i   \otimes_{b} \boldsymbol{\cal E}_i^*]\,\bvec((b\,\omega^{o*})^T\}\Big\}+\bvec(R_v^T)
\end{align}
Considering (\ref{eq:calBi}), matrix ${\cal F}$ can be written as:
\small
\begin{align}
{\cal F}&= \E\Big \{ \big[(I-{\cal M} \boldsymbol{\cal R}_i)^T (\boldsymbol{\cal A}_i+\boldsymbol{\cal E}_i) \big] \nonumber \\
&\hspace{2cm}  \otimes_{b} \big[(\boldsymbol{\cal A}_i+\boldsymbol{\cal E}^{*T}_i) (I-{\cal M} \boldsymbol{\cal R}_i) \big]\Big\} \nonumber \\
&= \E \Big \{ \big[(I-{\cal M} \boldsymbol{\cal R}_i)^T   \otimes_{b} (I-{\cal M} \boldsymbol{\cal R}_i) \big] \nonumber
\\
&\hspace{2cm} \times \big[(\boldsymbol{\cal A}_i+\boldsymbol{\cal E}_i)   \otimes_{b} (\boldsymbol{\cal A}_i+ \boldsymbol{\cal E}^{*T}_i) \big] \Big \}
\label{eq:calF-identity-3}
\end{align}
\normalsize
Since the entries of matrix $\boldsymbol{\cal R}_i$, which are defined in terms of the regression data $\u_{k,i}$, are independent of the entries of matrices
$\boldsymbol{\cal A}_i$ and $\boldsymbol{\cal E}_i$, i.e.,  $\a_{\ell,k}(i)$ and $\e_{\ell,k}(i)$, matrix ${\cal F}$ in (\ref{eq:calF-identity-3}) can be written more compactly as:
\begin{align}
{\cal F}= \bar{\cal F}\,{\cal D}
\label{eq:calF}
\end{align}
where
\begin{align}
&\bar{\cal F} \triangleq \E\left[(I-{\cal M} \boldsymbol{\cal R}_i)^T   \otimes_{b} (I-{\cal M} \boldsymbol{\cal R}_i) \right]
\label{eq.bar-cal-F}\\
&{\cal D} \triangleq \E[\boldsymbol{\cal D}_i]=\E \left[(\boldsymbol{\cal A}_i+\boldsymbol{\cal E}_i)   \otimes_{b} (\boldsymbol{\cal A}_i+ \boldsymbol{\cal E}^{*T}_i) \right]
\label{eq.cal-D}
\end{align}
We can find an expression for  $\bar{\cal F}$ if we assume that the regression data $\u_{k,i}$ are circular Gaussian---see equation (\ref{eq:bar-F-Gussian-2}) and Appendix \ref{apex.:four_order_moment_term_derivation}, where $e_k$ is a unit basis vector in ${\amsbb R}^N$ with entry one at position $k$, $r_k=\vec(R_{u,k})$, $\beta=2$ for real-valued data and $\beta=1$ for complex-valued data. A simplified  expression can be found to compute $\bar{\cal F}$ without using the Gaussian assumption on the regression data provided that the following condition holds.

\begin{assump}
\label{ass:small-step-size}
The channel estimation errors over the network are small enough such that the adaptation step-sizes in (\ref{eq.:mu_range}) can be chosen sufficiently small.
\end{assump}

\par \noindent
In cases where the distribution of the regression data is unknown, under Assumption \ref{ass:small-step-size}, the contributing terms  depending on $\mu^2_k$ can be neglected and as a result  $\bar {\cal F}$ can approximated by
\setcounter{equation}{80}
\begin{align}
\bar{\cal F}\approx \left[(I-{\cal M} {\cal R})^T   \otimes_{b} (I-{\cal M} {\cal R}) \right]
\label{eq.:ApproximateF}
\end{align}
In Appendix \ref{Appdix. CompuationOfD}, we show how to obtain the matrix ${\cal D}$ in (\ref{eq.cal-D}) needed for computing ${\cal F}$ in (\ref{eq:calF}). To evaluate $\gamma$, we use the following relations, which are also established in Appendix \ref{Appdix. CompuationOfD}:
\small
\begin{align}
&\E\big[\boldsymbol{\cal E}_i^T   \otimes_{b} \boldsymbol{\cal E}_i^* \big]=\E\big[\big(\boldsymbol{E}^T_i \otimes \boldsymbol{E}^{*}_i\big)\big] \otimes I_{M^2}
\label{eq.:t1} \\
&\E\big[(\boldsymbol{\cal A}_i+\boldsymbol{\cal E}_i)^T \otimes_{b} (\boldsymbol{\cal A}^T_i+\boldsymbol{\cal E}^{*}_i)\big] \nonumber \\
&=\Big(\E\big[(\boldsymbol{A}_i \otimes \boldsymbol{A}_i)^T\big]+\E\big[(\boldsymbol{A}^T_i \otimes \boldsymbol{E}^{*}_i)\big] \nonumber \\
&\hspace{1cm}+\E\big[(\boldsymbol{E}_i \otimes \boldsymbol{A}_i)^T ]+\E\big[\boldsymbol{E}^T_i \otimes \boldsymbol{E}^*_i \big] \Big) \otimes I_{M^2} \label{eq.:t2} \\
&\E\big[\boldsymbol{\cal B}_i   \otimes_{b} \boldsymbol{\cal E}_i^*]=\Big\{\big(\E\big[ (\boldsymbol{A}_i\otimes \boldsymbol{E}_i)^*]
+\E\big[\boldsymbol{E}^T_i\otimes \boldsymbol{E}_i^*]\big) \otimes I_{M^2} \Big\} \nonumber \\
&\hspace{2cm}\times \Big\{(I_{MN}-{\cal M}{\cal R})\otimes_b I_{MN} \Big\} \label{eq.:t3}
\end{align}
\normalsize
To obtain mean-square error (MSE) steady state expressions for the network, we let $i$ go to infinity and use expression (\ref{variance_relation_3}) to write:
\begin{align}
\lim_{i \rightarrow \infty}\E\|\tilde \w_i\|^2_{(I-{\cal F})\sigma}=\gamma^T\sigma
\label{eq.tild-wi-infinity}
\end{align}
Since we are free to choose $\Sigma$ and hence $\sigma$, we choose $(I-{\cal F})\sigma=\bvec(\Omega)$, where $\Omega$ is another arbitrary positive semidefinite matrix. Doing so, we arrive at:
\begin{align}
\lim_{i \rightarrow \infty}\E\|\tilde \w_i\|^2_{\Omega}=\gamma^T (I-{\cal F})^{-1} \bvec(\Omega)
\label{eq.tild-wi-infinity2}
\end{align}
Recall from (\ref{eq:tilde w_def}) that each sub-vector of ${\tilde \w}_i$ corresponds to the estimation error at a particular node, for instance, ${\tilde \w}_{k,i}$ is the estimation error at node $k$. Therefore, using (\ref{eq.tild-wi-infinity2}), the MSD at node $k$, denoted by $\eta_k$, can be computed by choosing $\Omega=\{\diag( e_k)\otimes I\}$, i.e.:
\begin{align}
\eta_k&=\lim_{i \rightarrow \infty}\E\|\tilde \w_{k,i}\|^2=\lim_{i \rightarrow \infty}\E\|\tilde \w_i\|^2_{\{\diag( e_k)\otimes I\}} \nonumber \\
&=\gamma^T (I-{\cal F})^{-1}\bvec(\diag(e_k)\otimes I_M)
\label{eq.:nodes_msd}
\end{align}
The network MSD, denoted by $\eta$, is then defined as:
\begin{equation}
\eta=\lim_{i \rightarrow \infty} \frac{1}{N} \sum_{k=1}^N \E\|\tilde \w_{k,i}\|^2
\end{equation}
which it can be also computed from (\ref{eq.tild-wi-infinity2}) by using $\Omega=\frac{1}{N} I_{MN}$. This leads to:
\small
\begin{align}
\eta=\lim_{i \rightarrow \infty} \frac{1}{N}\,\E\|\tilde \w_i\|^2=\frac{1}{N} \gamma^T (I-{\cal F})^{-1}\bvec(I_{M N})
\label{eq.:net_msd}
\end{align}
\normalsize
In (\ref{eq.:nodes_msd}) and (\ref{eq.:net_msd}), we assume that $(I-{\cal F})$ is invertible.  In what follows, we find conditions under which this assumption is satisfied. Using the properties of the Kronecker product and the sub-multiplicative property of norms, we can write:
\begin{align}
\rho({\cal F})&\leq  \| \bar{\cal F} {\cal D} \|_{b,\infty}  \leq \big\| \bar{\cal F} \big\|_{b,\infty} \|{\cal D}\|_{b,\infty}
   \label{eq.proof-calFstability-2}
\end{align}
We next show that $\bar{\cal F}$ from (\ref{eq.:ApproximateF}) is a block diagonal Hermitian matrix with block size $NM^2\times NM^2$. To this end, we note that $I-{\cal M} {\cal R}$ is a block diagonal matrix with block size  $M\times M$ and then use  (\ref{eq.:ApproximateF}) to obtain:
\begin{align}
\bar{\cal F}&=\diag\Big\{(I-\mu_1 R_{u,1})^T \otimes (I-{\cal M} {\cal R}),\nonumber \\
&\hspace{1cm}\cdots, (I-\mu_N R_{u,N})^T \otimes (I-{\cal M} {\cal R}) \Big \}
\end{align}
Moreover, $\bar{\cal F}$ is Hermitian because considering ${\cal R}={\cal R}^*$, ${\cal M}={\cal M}^T$, ${\cal R}{\cal M}={\cal M}{\cal R}$, we will have
\begin{align}
\bar{\cal F}^*& =\big((I-{\cal M} {\cal R})^T\big)^* \otimes_b (I-{\cal M} {\cal R})^*\nonumber \\
&= (I-{\cal M} {\cal R})^T \otimes_b (I-{\cal M} {\cal R})=\bar{\cal F}
\end{align}
Now we can use the following lemma to bound the spectral radius of matrix $\cal F$ in (\ref{eq.proof-calFstability-2}).
\newtheorem{lem}{Lemma}
\begin{lem}
\label{lem.:lemma1}
Consider an $N \times N$ block diagonal Hermitian
matrix $Y = \diag\{Y_1, Y_2,\cdots ,Y_N\}$, where each block $Y_k$ is of size $M \times M$ and Hermitian. Then it holds that\cite{sayed2012diffusion}:
\be
 \|Y\|_{b,\infty}=\max_{1 \leq k \leq N} \rho(Y_k)=\rho(Y)
\ee
\end{lem}
\par \noindent
According to this lemma, since $\bar{\cal F}$ is block diagonal Hermitian, we can substitute its block maximum norm on the right hand side of relation (\ref{eq.proof-calFstability-2}) with its spectral radius and obtain:
\begin{align}
\rho({\cal F})& \leq \rho \Big( (I-{\cal M} {\cal R})^T  \otimes_{b} (I-{\cal M} {\cal R}) \Big) \, \|{\cal D}\|_{b,\infty}
   \nonumber \\
    & = \rho^2 (I-{\cal M} {\cal R}) \; \|{\cal D}\|_{b,\infty}
    \label{eq.proof-calFstability-5}
\end{align}
We then deduce that $\rho({\cal F})<1$ if:
\be
0<\rho(I-{\cal M} {\cal R})<\frac{1}{\sqrt{\|{\cal D}\|_{b,\infty}}}
\ee
Since $I-{\cal M} {\cal R}$ is a block-diagonal matrix, this condition will be satisfied for small step-sizes that also satisfy: \begin{equation}
\frac{1-\frac{1}{\sqrt{\|{\cal D}\|_{b,\infty}}}}{\lambda_{\max}(R_{u,k})}<\mu_k < \frac{1+\frac{1}{\sqrt{\|{\cal D}\|_{b,\infty}}}}{\lambda_{\max}(R_{u,k})}
\label{eq.:mu_range_mean_square}
\end{equation}
If the channel estimation error is small, then $\|{\cal E}\|_{b,\infty} \approx 0$ and ${\cal D} \approx {\cal A}   \otimes_{b} {\cal A}$. Subsequently, $\|{\cal D}\|_{b,\infty} \approx 1$ and  this mean-square stability condition reduces to \mbox{$0<\mu_k< \frac{2}{\lambda_{\max}(R_{u,k})}$} which is the mean-square stability range of diffusion LMS over ideal communication links\cite{sayed2012diffusion}.

\subsection{Mean-Square Transient Behavior}
In this part, we derive expressions to characterize the mean-square convergence behavior of the diffusion algorithms over wireless networks with fading channels and noisy communication links. To derive these expressions, it is assumed that each node knows the CSI of its neighbors, and $\boldsymbol{E}_i=0$ for all $i$. We then use (\ref{variance_relation_1}) and consider $ \w_{k,-1}=0,\; \forall k \in\{1,\cdots,N\}$ to arrive at:
\begin{equation}
\E\|\tilde \w_{i}\|^2_{\sigma}=\| w^o\|^2_{{\cal F}^{i+1}{\sigma}}+ {\bar \gamma}^T\sum_{j=0}^i {\cal F}^j\sigma
\label{eq.:transient_recursion2}
\end{equation}
where
\small
\begin{align}
\bar{\gamma} &\triangleq  \E\big[\boldsymbol{\cal A}_i^T \otimes_b \boldsymbol{\cal A}_i\big] \bvec({\cal M} {\cal P}^T {\cal M})+\bvec(\bar {R}_v^T)\\
\bar{\cal R}_v & \triangleq     \diag\big\{{\bar R}_{v,1}\cdots, {\bar R}_{v,N}\big\}\\
{\bar R}_{v,k} & \triangleq  \E[\v^{(\psi)}_{k,i} \v^{(\psi)*}_{k,i}]=\sum_{\ell \in {\cal N}_{k} \backslash \{k\}} \E \big[\a^2_{\ell,k}(i) \, |\g_{\ell,k}(i)|^2\big] R^{(\psi)}_{v,\ell k}
\end{align}
\normalsize
Under this condition, and since $\boldsymbol{E}_i=0$,  ${\cal F}$ can be expressed as:
\vspace{-2mm}
\begin{align}
{\cal F} \approx \bar{\cal F} \; \E\big[\boldsymbol{\cal A}_i^T \otimes_b \boldsymbol{\cal A}_i\big]
\label{eq:bar-calF}
\end{align}
Writing (\ref{eq.:transient_recursion2}) for $i-1$ and computing $\E\|\tilde \w_{i}\|^2_{\sigma}-\E\|\tilde \w_{i-1}\|^2_{\sigma}$ leads to:
\begin{equation}
\E\|\tilde \w_{i}\|^2_{\sigma}=\E\|\tilde \w_{i-1}\|^2_{\sigma}+\| w^o\|^2_{{\cal F}^i(I-{\cal F}){\sigma}}+ \bar{\gamma}^T {\cal F}^i\sigma
\label{eq.:transient_recursion3}
\end{equation}
By replacing $\sigma$  with $\sigma_{\msd_k}=\diag( e_k)\otimes I_M$ and $\sigma_{\emse_k}=\diag( e_k)\otimes R_{u,k}$, we arrive at two recursions for the evolution of the MSD and EMSE over time:
\begin{align}
&\eta_k(i)=\eta_k(i-1)-\|w^o\|_{{\cal F}^i(I-{\cal F}){\sigma}_{\msd_k}}+\bar{\gamma}^T {\cal F}^i \sigma_{\msd_k} \label{eq.:msd-transient_state-k}\\
&\zeta_k(i)=\zeta_k(i-1)-\|w^o\|_{{\cal F}^i(I-{\cal F}){\sigma}_{\emse_k}}+\bar{\gamma}^T {\cal F}^i \sigma_{\emse_k}
\label{eq.:emse-transient_state-k}
\end{align}
We can find the learning curves of the network MSD and EMSE either by averaging the nodes learning curves (\ref{eq.:msd-transient_state-k}) and (\ref{eq.:emse-transient_state-k}), or by, respectively, substituting the following two values for $\sigma$ in recursion (\ref{eq.:transient_recursion3}):
\begin{align}
&\sigma_{\msd}=\frac{1}{N} \bvec(I_{MN}) \label{eq.:sigma-msd-network}\\
&\sigma_{\emse}=\frac{1}{N} \bvec\big(\diag\{R_{u,1},\cdots,R_{u,N}\}\big)
\label{eq.:sigma-emse-network}
\end{align}
%

\section{Numerical Results}
\label{sec.:DiffLmsFadingResults}
\begin{figure}[!h]
\centering
\includegraphics[width=6cm ,height=5cm]{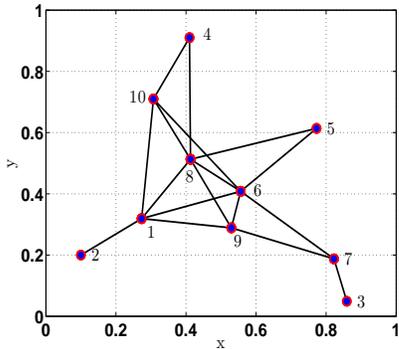}
\caption{\footnotesize{This graph shows the topology of the wireless network at the start-up time $i=0$, where two nodes are connected if their distance is less than their transmission range, $r_o=0.4$. }}
\label{fig.:Diffusion-topology}
\end{figure}

\begin{figure}
\centering
\includegraphics[width=6.8cm ,height=5.8cm]{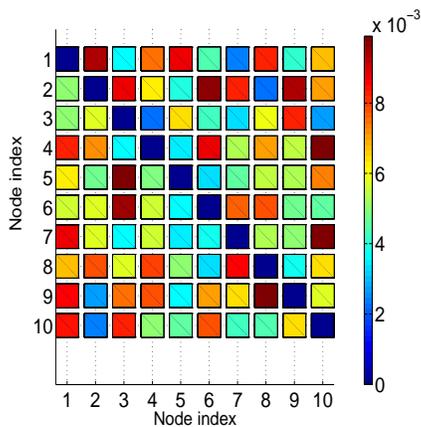}
\caption{\footnotesize{Power of communication noise
$\v_{\ell k,i}^{(\psi)}$ over the network. }}
\label{fig.:CommNoisePowerDiff}
\end{figure}

\begin{figure}
\centering
\includegraphics[width=5.4cm ,height=4.5cm]{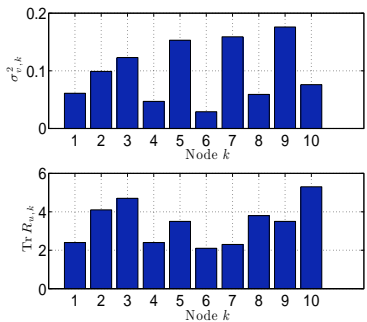}
\caption{\footnotesize{Network energy profile.}}
\label{fig.:network-energy-profile}
\end{figure}

In this section, we present computer experiments to illustrate the performance of the ATC diffusion strategy (\ref{alg.:atc-fadingLinkStep1})--(\ref{alg.:atc-fadingLinkStep2}) in the estimation of  the unknown parameter vector $w^o=2[1+j1,\, -1+j1]^T$ over time-varying wireless channels. We consider a network with $N=10$ nodes, which are randomly spread over a unit square area $(x,y)\in [0,\; 1] \times [0,\; 1]$, as shown Fig. \ref{fig.:Diffusion-topology}. We choose the transmit power of $P_{t}=1$, nominal transmission range of $r_{o}=0.4$ and the path-loss exponents $\alpha=3.2$.
For each node $k \in \{1,2,\cdots,N\}$,  we set $\mu_k=0.01$ and $\w_{k,-1}=0$. We adopt zero-mean Gaussian random distributions to generate $\v_{k}(i)$, $\v^{(\psi)}_{\ell k,i}$ and $\u_{k,i}$.  The distribution of the communication noise power over the spatial domain is illustrated in Fig. \ref{fig.:CommNoisePowerDiff}. The regression data $\u_{k,i}$ have covariance matrices of the form $R_{u,k}=\sigma^2_{u,k}I_M$. The trace of the regression data, $\Tr(R_{u,k})$, and the variances of measurement noise, $\sigma^2_{v,k}$, are illustrated in Fig. \ref{fig.:network-energy-profile}.

The exchanged data between nodes experience distortion characterized by (\ref{eq.:receivedWeight}).  At time $i$, the link between nodes
$\ell$ and $k$ fails with probability $1-p_{\ell,k}$.
%
We obtain $\gamma_{\ell,k}$ using the relative-degree combination rule \cite{cattivelli2010diffusion,sayed2012diffusion}, i.e.,
\begin{equation}
\gamma_{\ell,k}= \left\{ \begin{array}{l l}
\frac{|{\cal N}_{\ell}|}{\sum_{m \in {\cal N}_{k}} |{\cal N}_{m}|}, & \textrm{if} \, {\ell \in {\cal N}_{k}} \\
0, & \textrm {otherwise} \\
\end{array}
\right .
\label{eq.:relative-degree}
\end{equation}
and update $\A_i$ it at each time $i$ according to the introduced combination rule (\ref{eq.:Ai}).

\begin{figure}
\centering
\includegraphics[width=9cm ,height=7cm]{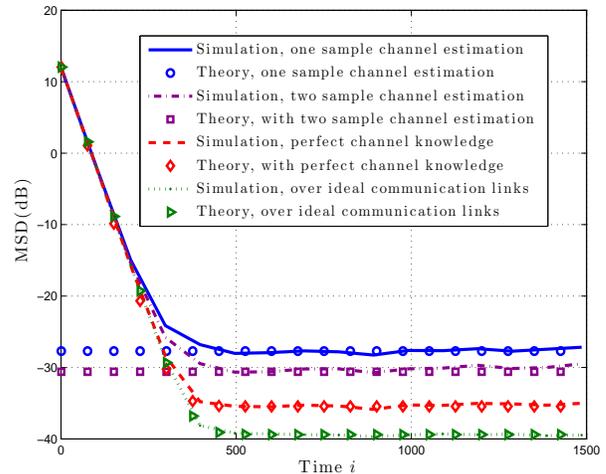}
\caption{Learning curves of the network in terms of MSD and EMSE.}
\label{fig.:trans_state_MSD}
\end{figure}

\begin{figure}
\centering
\includegraphics[width=9cm ,height=7cm]{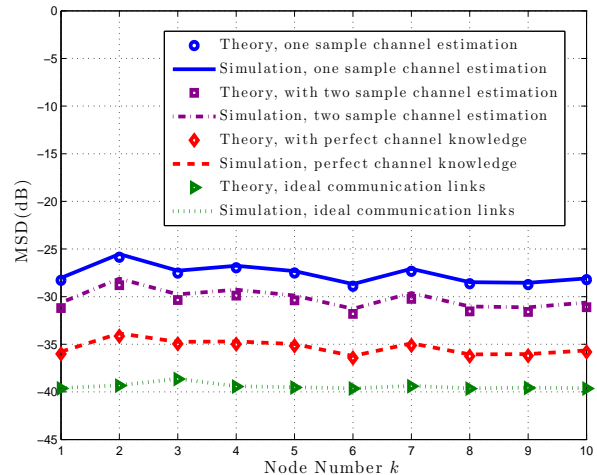}
\caption{Steady-state MSD over the network.}
\label{fig.:steady_state_MSD}
\end{figure}

Figures \ref{fig.:trans_state_MSD} and \ref{fig.:steady_state_MSD} show the network MSD in transient and steady-sate regimes, where the simulation curves  are obtained from the average of $500$ independent runs. In these figures, we compare the performance of the proposed ATC diffusion algorithm over wireless channels for different CSI cases at the receiving nodes. In particular, we examine the performance of the algorithm with perfect CSI, where each node $k$ knows the CSI of all its neighbors. We also consider scenarios where nodes do not have access to the CSI of their neighbors and obtain this information using one and two samples pilot data. For reference, we also illustrate the performance of ATC diffusion over ideal  communication links in which the communication links between nodes are error-free, i.e., for each node $k$,  $\psi_{\ell k,i}=\psi_{\ell,i}$ for all $i$.

The best performance in these experiments belongs to the diffusion strategy that runs over network with ideal communication links. As expected, the diffusion strategy with perfect CSI knowledge outperforms diffusion strategy with channel estimation using one or two samples pilot data, respectively, by 5dB and 7dB. In particular, the steady-sate mean-square performance of the algorithm improves almost by 2dB for an additional sample of pilot data used for channel estimation. Therefore, if the wireless channels are slowly-varying, by using a larger number of pilot data, it is possible to approach the performance of the diffusion strategy algorithm with perfect CSI.

We have also produced a transient  MSD curve using standard diffusion LMS \cite{cattivelli2010diffusion}, under similar fading conditions and noise. The results showed that the network MSD grows unbounded \mbox{(i.e.,  error  $\rightarrow \infty$}).  This problem can be justified using the fact that  some nodes, in the combination step,  use severely distorted data from neighbors with bad channel conditions and low SNR. Consequently a large error is introduced into their updated intermediate estimates, which then will propagate into the network in the following iterations and cause catastrophic network failure.

%
\begin{figure}
\centering
\includegraphics[width=9cm ,height=7cm]{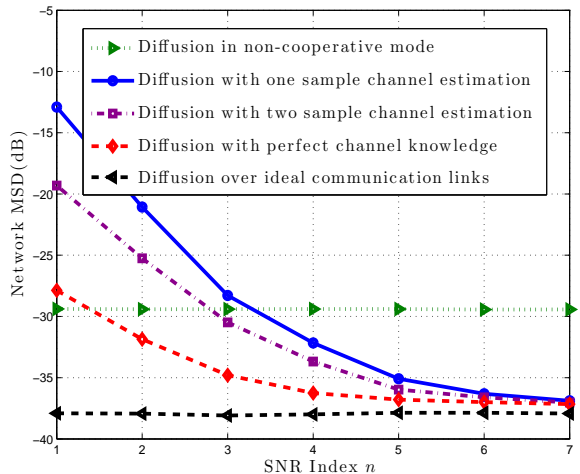}
\caption{The network performance comparison with non-cooperative diffusion LMS and with diffusion LMS over ideal communication links.}
\label{fig.:MsdVsSnr}
\end{figure}

In Fig. \ref{fig.:MsdVsSnr}, we compare the performance of diffusion strategies for different ranges of SNR over the network. We also make some comparisons between the cooperative and non-cooperative networks where in the latter case the network runs a stand-alone LMS filter at each node, which is equivalent to running the diffusion strategy with $\A_i=I$. In Fig. \ref{fig.:MsdVsSnr}, the SNR index $n \in \{1,2,\cdots,7\}$ over the $x$-axis refers to the $n$-th network SNR distribution, as obtained by uniformly scaling up the initial SNR distribution over the network by 5dB for each increment in the integer $n$, as represented by $\text{SNR}_n = \text{SNR}_{\text{ini}} +5n\,\text{(dB)}$,
where $\text{SNR}_{\text{ini}}$ are the SNR of the connected nodes illustrated in Fig. \ref{fig.:Diffusion-topology}, and are obtained from uniformly distributed random variables in the range between $[5\; 10]$dB.

As shown in Fig. \ref{fig.:MsdVsSnr}, the performance of non-cooperative adaptation and diffusion LMS with ideal communication links remains invariant with changes in the SNR values. This is expected since the performance of the diffusion LMS in these cases is not affected by the communication noise, $\v^{(\psi)}_{\ell k,i}$ and $\v^{(y)}_{\ell k,i}$.
In comparison, the performance of the modified diffusion strategy over wireless links depends on the CSI. As the knowledge about the network CSI increases, the performance improves. From this result, we observe that at low SNR the performance discrepancies between diffusion with perfect CSI and diffusion with channel estimation is larger compared to high SNR scenarios. This difference in performance can be reduced by using more pilot data to estimate the channel coefficients in each time slot. In addition, at very low SNR, we see that the non-cooperative case outperforms the modified diffusion strategy. This result suggests that in wireless networks with high levels of communication noise at all nodes (e.g., when the nodes transmit power is very low), to maintain a satisfactory performance level the network must switch to the non-cooperative mode. This also suggests that if the transmit power of some nodes is below some threshold value, these nodes should go to a sleep mode in order to avoid error propagation over the network.

\section{Conclusion}
\label{sec.:conclusion}
We extended the application of diffusion LMS strategies to sensor networks with time-varying fading wireless channels.
We analyzed the convergence behavior of the modified diffusion LMS algorithms, and established conditions under which the algorithms converge and remain stable in the mean and mean-square error sense. The analysis revealed that the performance of the diffusion strategies highly depend on the level of CSI knowledge and the level of communication noise power over the network. In particular, when the CSI are known, the modified diffusion algorithms are asymptotically unbiased and converge in the slow adaptation regime. In contrast, the parameter estimates will become biased when the CSI are obtained through pilot-aided channel estimation. Nevertheless, the size of the bias can be made small by increasing the number of pilot symbols or increasing the link SNR.

\clearpage
\pagestyle{empty}
\newpage
\appendices
\section{Computation of $R_{v,k}$}
\label{Appdix:Computation-of-Rvk}
To obtain $R_{v,k}$ in (\ref{eq:Rvlk}), we need to compute the expectation

\be
\E\left[\a^2_{\ell,k}(i) \,|{\hat \g}_{\ell,k}(i)|^2\right]=
\E \left[\frac{\a^2_{\ell,k}(i)}{\frac{P_t}{r^\alpha_{\ell,k}}| \hat{\h}_{\ell,k}(i)|^2 } \right]
\label{eq.a2g2}
\ee
for $\ell \in {\cal N}_k \backslash k$. For the case $\ell= k$, we have $R_{v,\ell k}=0$ and hence the expectation of $[\a^2_{\ell,k}(i) \,|{\hat \g}_{\ell,k}(i)|^2] R_{v,\ell k}$ in (\ref{eq:Rvlk}) is zero.
For $\ell \neq  k$, we proceed as follows. Since the joint probability distribution function of the numerator and denominator in (\ref{eq.a2g2}) is unknown, the expectation can be approximated using one of two ways. In the first method, we can resort to computer simulations. In the second method, we can resort to a Taylor series approximation as follows. We introduce the real-valued auxiliary variable $\x=\a^2_{\ell,k}(i)$. Considering the combination rule (\ref{eq.:Ai}), the expectation of $\x$  when $\ell \neq k$ will be:

\be
\E[\x]=\gamma^2_{\ell k} p_{\ell,k}
\ee
\setcounter{equation}{125}
\begin{figure*}[!t]
\begin{align}
\bvec\big(\E\big[\boldsymbol{\cal R}_i {\cal Q}' & \boldsymbol{\cal R}_i \big]\big)=
\Bigg \{ ({\cal R}^T {\otimes}_b {\cal R}) +\sum_{k=1}^N \Big[ \diag(\big(\vec(\diag(e_k))) \Big]\otimes \Big[ (\beta-1) (R^T_{k,u} \otimes R_{k,u})+r_k r^*_k \Big] \Bigg \} ({\cal M} \otimes_b {\cal M}) \bvec({\cal Q})
\label{eq:bvec-RQ'R}
\end{align}
\hrule
\end{figure*}
To compute the variance and expectation of the denominator in (\ref{eq.a2g2}), we let the exponential distribution function $f_{\y}(y)$ with parameter
$\lambda$ given by (\ref{eq.:LambdaForh}) denote the pdf of $\y= |{\hat \h}_{\ell,k}(i)|^2$, i.e.,
\setcounter{equation}{109}

\be
f_{\y}(y)=\lambda_{\ell,k} e^{-\lambda_{\ell,k} y}, \for y \in [0,\, \infty)
\label{eq.:fx}
\ee
We also let $f^{(t)}_{\y}(y)$ represent the pdf of $\y$ for  $y \in [\nu_{\ell,k},\, \infty)$.
It can be verified that $f^{(t)}_{\y}(y)$ represents a truncated exponential distribution and is given by:

\begin{align}
f^{(t)}_{\y}(y)=\lambda_{\ell,k} e^{-\lambda_{\ell,k} (y-\nu_{\ell,k})}, \for y \in [\nu_{\ell,k},\, \infty)
\label{eq.:hat-fx}
\end{align}
If we now define

\be
\z=\frac{P_t}{r^\alpha_{\ell,k}} \y
\ee
Then, the pdf of $\z$ can be computed as \cite{Garcia1994probability}:
\be
f_{\z}(z)=\Big|\frac{dy}{dz}\Big| f^{(t)}_{\y}(g^{-1}(z))
\label{eq.:fZ1}
\ee
where
\begin{align}
\frac{d y}{d z}=\frac{r^\alpha_{\ell,k}}{P_t}\; \text{and}\; g^{-1}(z)=\frac{r^\alpha_{\ell,k}}{P_t} z
\label{eq.:g-inverse}
\end{align}
Therefore,
\be
f_{\z}(z)=\frac{r^\alpha_{\ell,k}}{P_t} \lambda_{\ell,k} e^{-\lambda_{\ell,k} (\frac{r^\alpha_{\ell,k}}{P_t} z-\nu_{\ell,k})}, \for z \in \Big[\frac{P_t}{r^\alpha_{\ell,k}} \nu_{\ell,k},\, \infty \Big)
\label{eq.:fZ-final}
\ee
Using this distribution the mean and  variance of $\z$ will be \cite{Garcia1994probability}:
\begin{align}
&\E[\z]=\frac{P_t}{r^\alpha_{\ell,k}}\Big(\frac{1}{\lambda_{\ell,k}}+\nu_{\ell,k}\Big)  \\
&\var(\z)=\Big(\frac{P_t}{r^\alpha_{\ell,k}\lambda_{\ell,k}}\Big)^2
\end{align}
We can now proceed to approximate the expectation (\ref{eq.a2g2}) by defining
\be
f(\x,\z)=\frac{\x}{\z}
\ee
and employing a second order Taylor series expansion  to write:
\be
\E[f(\x,\z)] \approx \frac{\E[\x]}{\E[\z]}-\frac{1}{(\E[\z])^2} \text{cov}(\x,\z)+ \frac{\E[\x]}{(\E[\z])^3}\var(\z)
\label{eq.:Efxz1}
\ee
Substituting, $\E[\x]$, $\E[\z]$, cov$(\x,\z)$ and var$(\z)$ into (\ref{eq.:Efxz1}), we then arrive at:
\begin{align}
&\E[f(\x,\z)] \approx \E \left[\a^2_{\ell,k}(i) \,|{\hat \g}_{\ell,k}(i)|^2\right] \nonumber \\
& \approx \gamma^2_{\ell k} p_{\ell,k}\Big(\frac{1}{\frac{P_t}{r^\alpha_{\ell,k}}(\frac{1}{\lambda_{\ell,k}}+\nu_{\ell,k})} -\frac{\nu_{\ell,k}}{\frac{P_t}{r^\alpha_{\ell,k}}(\frac{1}{\lambda_{\ell,k}}+\nu_{\ell,k})^2}
\nonumber \\
&\hspace{2cm} +\frac{1} {\frac{P_t}{r^\alpha_{\ell,k}}\lambda_{\ell,k}^2(\frac{1}{\lambda_{\ell,k}}+\nu_{\ell,k})^3}\Big)
\label{eq.:excationOfa_lk2v_lk2}
\end{align}

\section{Derivation of (\ref{eq:bar-F-Gussian-2})}
\label{apex.:four_order_moment_term_derivation}
First, we note that when $\u_{k,i}$ are zero mean circular complex-valued Gaussian random vectors and i.i.d. over time, then for any Hermitian matrix $\Gamma$ of compatible dimensions it holds that \cite{sayed2008}:
\begin{equation}
\E[\u^{\ast}_{k,i} \u _{k,i}\Gamma  \u ^*_{k,i} \u_{k,i}]=\beta(R_{u,k}\Gamma R_{u,k})+R_{u,k} \Tr(\Gamma R_{u,k})
\label{fourth_order_regression}
\end{equation}
where $\beta=1$ for complex regressors and $\beta=2$  when the regressors are real. Using (\ref{fourth_order_regression}) and spatial independence of the regression data we have
\begin{align}
\E[\u^{\ast}_{k,i} \u _{k,i}\Gamma  \u ^{\ast}_{\ell,i} \u_{\ell,i}]&=R_{u,k}\Gamma R_{u,\ell}+\delta_{k \ell}(\beta-1) R_{u,k}\Gamma R_{u,k} \nonumber \\
&\qquad +\delta_{k \ell} R_{u,k} \Tr(\Gamma R_{u,k})
\label{fourth_order_regression_extention}
\end{align}
where $\delta_{k \ell}$ is the Dirac delta sequence. To compute $\bar{\cal F}$, we first introduce
\be
\boldsymbol{\cal L}_i=(I-{\cal M}\boldsymbol{\cal R}_i) {\cal Q} (I-{\cal M}\boldsymbol{\cal R}_i)
\label{eq.:cal-L}
\ee
where ${\cal Q}$ is an arbitrary deterministic Hermitian matrix.
We now note that
\begin{align}
\bvec\left(\E[\boldsymbol{\cal L}_i] \right) &\numrel{=}{label-1} \E \left[(I-{\cal M} \boldsymbol{\cal R}_i)^T \otimes_b (I-{\cal M}\boldsymbol{\cal R}_i)\right] \bvec({\cal Q}) \nonumber \\
&\numrel{=}{label-2}\bar{\cal F}\; \bvec({\cal Q})
\label{eq:cal-L2}
\end{align}
where (\ref{label-2}) obtained by comparing the expectation term on the right hand side of (\ref{label-1}) with definition (\ref{eq.bar-cal-F}).
We proceed by taking expectation of both sides of $(\ref{eq.:cal-L})$, i.e.,
\be
\E[\boldsymbol{\cal L}_i]={\cal Q}-{\cal R} {\cal M}  {\cal Q}-{\cal Q} {\cal M} {\cal R}+\E\left[\boldsymbol{\cal R}_i {\cal M} {\cal Q} {\cal M}
\boldsymbol{\cal R}_i\right]
\label{eq.:E[cal-L]-1}
\ee
To compute the block vectorization of the last term on the right hand side of (\ref{eq.:E[cal-L]-1}), we introduce the block partitioned matrix ${\cal Q}'={\cal M} {\cal Q} {\cal M}$ with blocks ${\cal Q}'_{k\ell}$ and use (\ref{fourth_order_regression_extention}) to obtain  (\ref{eq:bvec-RQ'R}), where $r_k=\vec(R_{u,k})$.

Now, using (\ref{eq.:E[cal-L]-1}), we can write:
\setcounter{equation}{126}
\begin{align}
\bvec(\E[\boldsymbol{\cal L}_i])=&\big(I-I\otimes_b{\cal M}{\cal R}-{\cal R}^T {\cal M} \otimes_b I\big)\bvec(\cal{Q})\nonumber \\
&\hspace{1cm}+\bvec\big(\E\big[\boldsymbol{\cal R}_i {\cal Q}' \boldsymbol{\cal R}_i \big]\big)
\label{eq:bvec-E[cal-L]-1}
\end{align}
From (\ref{eq:cal-L2}), (\ref{eq:bvec-RQ'R}) and (\ref{eq:bvec-E[cal-L]-1}) and using the fact that
the real vector space of Hermitian matrices is isomorphic to ${\amsbb R}^{N^2\times 1}$ \cite{gallier2011geometric}, we arrive at (\ref{eq:bar-F-Gussian-2}).
\section{Computation of ${\cal D}$}
\label{Appdix. CompuationOfD}
We expand ${\cal D}=\E[\boldsymbol{\cal D}_i]$ in (\ref{eq.cal-D}) as:
\begin{align}
{\cal D}
=\Big\{\E[\boldsymbol{A}_i \otimes  \boldsymbol{A}_i]+\E[\boldsymbol{A}_i \otimes  \boldsymbol{E}_i^{*T}]+\E[\boldsymbol{E}_i \otimes \boldsymbol{A}_i] \nonumber \\
+\E[\boldsymbol{E}_i \otimes \boldsymbol{E}_i^{*T}] \Big\} \otimes I_{M^2}
\label{eq.:D-terms}
\end{align}
The $(r,\,z)$-th entry of $\E[\boldsymbol{A}_i \otimes  \boldsymbol{A}_i]$, denoted by $f_{r,z}$, is:
\begin{align}
f_{r,z}=\E[\a_{\ell,k}(i) &\a_{m,n}(i)]
\end{align}
where the relation between $(r,z)$ and $\ell,k)$ is:
\begin{align}
r=(\ell-1)N+m,  \; \text{and} \; z=(k-1)N+n   \label{eq.:s-entry}
\end{align}
When $k\neq n$, entries $\a_{\ell,k}(i)$ and $\a_{m,n}(i)$ come from different columns of $\boldsymbol{A}_i$ and are independent. Hence, in this case, we can write:
\begin{align}
f_{r,z}=\E\big[\a_{\ell,k}(i)\big] \, \E\big[\a_{m,n}(i)\big]
\end{align}
with
\begin{equation}
\E\big [\a_{j,q}(i)\big ]=\left\{ \begin{array}{l l}
1-\displaystyle \sum_{r \in {\cal N}_q \backslash q }\, p_{r q} \gamma_{r q}, & \textrm{if}\,\, j= q  \\
 p_{jq} \gamma_{jq}, &\textrm{otherwise} \\
\end{array}
\right .
\end{equation}
When $k=n$, the entries $\a_{\ell,k}(i)$ and $\a_{m,n}(i)$ come from the same column of $\boldsymbol{A}_i$ and may be dependent.
In this case, there are four possibilities:
\par \noindent
(1) if $\ell=m$ and $\ell \neq k$:
\begin{align}
f_{r,z}=\gamma_{\ell,k}^2 p_{\ell,k}
\end{align}
(2) if $\ell=m$ and $\ell =k$:
\begin{align}
&f_{r,z}=\E\Big[\Big(1-\sum_{\ell \in {\cal N}_k \backslash k} \a_{\ell,k}(i)\Big)\Big(1-\sum_{\ell \in {\cal N}_k \backslash k}\a_{\ell,k}(i)\Big)\Big]\\
&=1-2\sum_{\ell \in {\cal N}_k \backslash k} p_{\ell,k} (\gamma_{\ell,k}-\gamma^2_{\ell,k})-\sum_{\ell \in {\cal N}_k \backslash k}
p^2_{\ell,k}\gamma^2_{\ell,k} \nonumber \\
&\quad +{\sum_{(\ell \in {\cal N}_k \backslash k)} \sum_{(m \in {\cal N}_k \backslash k)}} p_{\ell,k} p_{m,k} \gamma_{\ell,k}\gamma_{m,k}
\end{align}
\par \noindent
(3) if $\ell \neq m$ and $\ell \neq k$ and $m \neq n$:
\begin{align}
f_{r,z}=\gamma_{\ell,k} \gamma_{m,n} p_{\ell,k} p_{m,n}
\end{align}
(4) if $\ell \neq m$ and $\ell = k$ and $m \neq n$:
\begin{align}
f_{r,z}&=\E\Big[ \big(1-\sum_{j \in {\cal N}\backslash k } \a_{j,k}(i)\big ) \a_{m,n}(i) \Big ] \nonumber  \\
&= \gamma_{m,n} p_{m,n}\Big (1 -\gamma_{m,n}+ \sum_{j \in {\cal N}_k \backslash\{k,m\}} \gamma_{j, k} p_{j,k}\Big)
\end{align}

The $(r,\,z)$-th entry of $\E[\boldsymbol{A}_i \otimes  \boldsymbol{E}_i^{*T}]$, denoted by $x_{r,z}$, can be expressed as:
\small
\begin{align}
&x_{r,z}=-\E\Big[\a_{\ell,k}(i)\, \a_{m,n}(i) {\hat \g}^*_{m,n}(i) \v^{(y)*}_{m,n}(i) \Big] \nonumber\\
&=-\E\Bigg[\a_{\ell,k}(i) \a_{m,n}(i) \frac{\sqrt{\frac{r^{\alpha}}{P_t}}{\h}_{m,n}(i) \v_{m,n}^{(y)*}(i)+\frac{r^{\alpha}}{P_t}|\v_{m,n}^{(y)}(i)|^2}{\big|{\h}_{m,n}(i)+\sqrt{\frac{r^{\alpha}}{P_t}}\v_{m,n}^{(y)}(i)\big|^2} \nonumber \\
& \hspace{1.5cm} \Big| \big|{\h}_{m,n}(i)+\sqrt{\frac{r^{\alpha}}{P_t}}\v_{m,n}^{(y)}(i)\big|^2 \geq  \nu_{m,n}
\Bigg]
\label{e:x_rz}
\end{align}
\normalsize
Likewise, the entries of $\E[\boldsymbol{E}_i \otimes \boldsymbol{A}_i]$ and $\E[\boldsymbol{E}_i \otimes \boldsymbol{E}_i^{*T}]$ can be expressed in terms of the combination weights, channel coefficients and the estimation error. We can follow the argument presented in Remark \ref{re:remark-1}  to show that the right hand side of (\ref{e:x_rz}) as well as the entries of $\E[\boldsymbol{E}_i \otimes \boldsymbol{A}_i]$ and $\E[\boldsymbol{E}_i \otimes \boldsymbol{E}_i^{*T}]$ are invariant with respect to time and have finite values.

\end{document}